\documentclass[st,twocolumn]{jpsj3}
\usepackage{txfonts}
\usepackage{amsmath}
\usepackage{color}

\title{Quantum annealing of pure and random Ising chains
coupled to a bosonic environment}

\author{Sei Suzuki$^1$\thanks{sei01@saitama-med.ac.jp}, Hiroki
Oshiyama$^2$, and Naokazu Shibata$^2$}
\inst{$^1$Department of Liberal Arts, Saitama Medical University, Moroyama, Saitama 350-0495, Japan \\
$^2$Department of Physics, Tohoku University, Sendai 980-8578, Japan}

\abst{We study quantum annealing of pure and random transverse Ising chains 
that are coupled to boson baths, using a novel
numerical method based on the combination of
the quasi-adiabatic propagator path
integral (QUAPI) and the matrix product state (MPS) formalisms.
We present numerical results on systems up to $64$ spins
and conclude that the baths disturb quantum annealing of
pure and random Ising chains.
The numerical method to compute the reduced density matrix
is described in detail.
}

\begin{document}
\maketitle

\section{Introduction}

Controlling the time evolution of a quantum many-body system has been
an important issue for decades in condensed matter physics, quantum
physics, statistical physics,
information science, and engineering.
Today's swell in the study of this issue stems partly from recent
advances in quantum computing technology. In particular, the appearance
of quantum annealing processor has stimulated research interests quite strongly.
Although the quantum annealing processor has not been as useful as a
conventional computer so far, it has attracted quite a lot of hopes that
the first practical quantum computer using quantum annealing should appear
soon.

One of the biggest problems associated with the quantum annealing processor 
is that no one has known the theoretical power of it so far. 
Quantum annealing was proposed originally by assuming an isolated interacting
spin system\cite{bib:KadowakiPRE1998,bib:FarhiArXiv2000}.
The power of quantum annealing in an isolated system has been
studied for the last two decades and clarified to some extent\cite{bib:SuzukiEPJST2015}.
The simplest model of quantum annealing with the transverse field
is apt to fail in solving typical problems such as 3-Satisfiability
because of a first-order quantum phase
transition and/or a localization phenomena of the wave function
\cite{bib:JoergPRL2010}.
In realistic situations, however, the system which quantum annealing
is applied to cannot be isolated from but must be inevitably open
to an environment.
This is true as for the current quantum annealing processor made by
D-Wave System Inc. It adopts superconducting flux qubits as physical
spins, which are coupled with an environment through a fluctuation in
the potential energy of a flux caused by the normal current flowing
across the Josephson junction.
In fact, the current quantum annealing processor shows properties that
are different from the ideal quantum annealing\cite{bib:BoixoNComm2016}.
Therefore it is inevitably important to consider 
an open system coupled to an environment in order to clarify the power 
of quantum annealing processors.

The study of time evolution in an open quantum system is significant
in the context different from quantum computation as well. The time evolution
of a closed (i.e., isolated) quantum many-body system with driving has been studied
intensively and a lot of theoretical and experimental progresses
have been attained in the last few decades. On the other hand,
as we lack analytical or numerical tools that are
useful to solve many-body problems with an environment, 
an open quantum many-body system has not been understood as much as
a closed system. 
An open system, however, is with no doubt as signifant as a closed system.
In fact, time evolution of an open system has started receiving
attentions independently of quantum computation.
Thus the interests in the time evolution of an open quantum many-body
system are growing in the wide range of communities.

In the present paper, we focus on the transverse Ising chain
coupled to bosonic baths. This model is standard and has been studied
in some earlier works. The single spin version of it was studied
extensively by Leggett et al.\cite{bib:LeggettRMP1987} thirty years ago.
An accurate numerical method, named the
quasiadiabatic propagator path integral (QUAPI), was developed to study the time evolution
of a single spin coupled to a bosonic bath by Makarov and Makri
in 1994\cite{bib:MakarovCPL1994}. After that, QUAPI was applied to study the 
Landau-Zener model coupled to an environment by Nalbach and
Thorwart\cite{bib:NalbachPRL2009} in 2009.
The time evolution of a driven many-spin system coupled to a bosonic
bath was first studied by Patan\`{e}
et al.\cite{bib:PatanePRL2008,bib:PatanePRB2009},
where modified Kibble-Zurek scalings were discussed.
The Kibble-Zurek scaling in the presence of 
baths has been also studied by Nalbach et al.\cite{bib:NalbachPRB2015} and
Dutta et al.\cite{bib:DuttaPRL2016}
The pure transverse Ising chain coupled to a bosonic chain
has been studied in the context of quantum annealing by
Smelyanskiy et al.\cite{bib:SmelyanskiyPRL2017} and 
Arceci et al.\cite{bib:Arceci2018} recently.
Quantum annealing of random transverse Ising models in an environment
has been studied in
Refs. \citen{bib:AminPRA2009,bib:AminPRA2015,bib:BoixoNComm2016,bib:KechedzhiPRX2016},
for the purpose to clarify the performance of D-Wave's quantum
annealing processors.

Most of these theoretical studies on many-spin systems
coupled to bosonic baths have assumed the so-called Born-Markov (BM) approximation
and/or an integrable spin system.
The BM approximation is expected to be valid for weak coupling limit between
the system and the environment. It however involves a drawback that the
approximation is uncontrollable, namely, one cannot improve the
accuacy of the approximation systematically.
We resort to neither the BM approximation nor the integrability
of a spin system. We 
instead develop a novel method that is based on the QUAPI and the
matrix product state (MPS) formalism.
Our method employs approximations in a controllable manner and
one can attain the exact result in an appropriate limit.
The accessble size of the system reaches $N\sim 10^2$.
Our method is
applicable to one-dimensional quantum systems with bosonic baths.
The present paper is devoted to describe this method and to present
some results obtained using it.

This paper is organizaed as follows. We introduce the
transverse Ising model coupled to bosonic baths in
Sec.~\ref{sec:Model}.
Some of the numerical results are
presented in Sec.~\ref{sec:Results}, where we mention how
an error after quantum annealing is influenced by
a bosonic environment.
The numerical method is descirbed in detail in Sec.~\ref{sec:Method}.
Performing the partial trace in the density operator
with respect to the bosonic degree of freedom of the baths,
we obtain the QUAPI formula for the reduced density matrix. 
We provide an expression for it in Sec. \ref{sec:QUAPI}.
In order to compute the time evolution of the reduced density
matrix in systems with more than 10 spins, we needs a trick
to avoid an exponential increase in the number of states.
We introduce the
MPS formalism to reduce the number of bases in Sec. \ref{sec:DMRG}.
We conclude the present paper in Sec. \ref{sec:Conclusion}

\section{Model}\label{sec:Model}

We consider the transverse Ising chain coupled to bosonic baths.
The Hamiltonian is composed of
$H_{\rm S}$, $H_{\rm B}$ and $H_{\rm int}$, representing
the spin system, the baths, and the interaction
between them, respectively:
\begin{equation}
 H(t) = H_{\rm S}(t) + H_{\rm B} + H_{\rm int} .
\end{equation}

The system is the transverse Ising chain that is written as
\begin{equation}
 H_{\rm S}(t) = - \sum_{j=1}^{N} J_j(t)\sigma_j^z\sigma_{j+1}^z
  - h(t)\sum_{j=1}^N \sigma_j^x ,
\end{equation}
where $\sigma_j^{\alpha}$ ($\alpha = x, z$) are the Pauli's matrices
at site $j$ and $N$ is the number of sites.
The spin-spin coupling constant and the transverse field are
denoted by $J_j(t)$ and $h(t)$, respectively. 
We focus on spatially uniform $h(t)$ for simplicity, though
it is possible to consider nonuniform $h(t)$.
It is noted that $J_j(t)$ and $h(t)$ may depend on time.
We assume the open boundary condition.
In standard quantum annealing, one considers $H_{\rm S}(t)$ such that
$h(0) \gg J_j(0)$ and $h(\tau) \ll J_j(\tau)$ for a given
runtime $\tau$, namely, the transverse field initially dominates
the Hamiltonian while the Hamiltonian coincides with
the Ising Hamiltonian finally.
The solution of the optimization to be solved is encoded
into the ground state of this Ising Hamiltonian.
If the system is isolated and the change speed of the
Hamiltonian is sufficiently slow, an adiabatic time evolution
from the simple ground state of $H_{\rm S}(0)$ to $H_{\rm S}(\tau)$
brings us the solution.

The baths are collections of harmonic oscillators whose
Hamiltonian is written as
\begin{equation}
 H_{\rm B} = \sum_{j=1}^N\sum_a \omega_{a}b_{ja}^{\dagger}b_{ja} ,
  \label{eq:H_B}
\end{equation}
where $a$ stands for the mode of harmonic oscillators. 
$b_{ja}$ and $b_{ja}^{\dagger}$ are the bosonic creation and
annihilation operators for the site $j$ and the mode $a$.
$\omega_{a}$ is the enregy of the harmonic oscillator in the
unit of $\hbar = 1$.
We assume that each site has own bath independently.

The interaction between the spin system and the environment
is given by
\begin{equation}
 H_{\rm int} = \sum_{j=1}^N\sigma_j^z\sum_a\lambda_{ja}
  \left(b_{ja}^{\dagger} + b_{ja}\right) ,
\label{eq:Hint}
\end{equation}
where $\lambda_{ja}$ determines the strength of the
interaction. This interaction Hamiltnoian
implies that each spin couples with own bath independently.
We note that one may in general consider coupling between
$\sigma_j^{\alpha}$ with $\alpha = x$ as well as $z$
and the bath. In fact, several previous studies
\cite{bib:PatanePRL2008,bib:PatanePRB2009,bib:NalbachPRB2015,
bib:DuttaPRL2016,bib:KeckNJP2017} have
considered coupling through $\sigma_j^{x}$,
since this assumption makes the model
a free fermionic model through the Jordan-Wigner transformation.
However, it has been know that the
coupling through $\sigma_j^z$ is
dominant over $\sigma_j^x$
\cite{bib:AverinPRB2000,bib:LantingPRB2011} in
the system of superconducting flux qubits of which the
D-Wave's quantum annealing processor is composed.
Hence, in the present paper, we focus our attention
on the coupling through $\sigma_j^z$ and consider
Eq. (\ref{eq:Hint}).

For the spectrum of the bosonic baths, we define the spectral
density as
\begin{equation}
 J(\omega) = \sum_a \lambda_{ja}^2\delta(\omega - \omega_{a}) ,
\label{eq:SpectralDensity1}
\end{equation}
where $\delta(\omega - \omega_{a})$ stands for the Dirac's delta
function. We then assume a continuous spectral density as
\begin{equation}
 J(\omega) = g^2 \omega^s e^{-\omega/\omega_{\rm c}} \theta(\omega),
\label{eq:SpectralDensity2}
\end{equation}
and $s = 1$, namely, the Ohmic spectral density
thoughout this paper for the sake of simplicity.
Here $g^2$ and $\omega_{\rm c}$ are the coupling constant,
between the spin system and the bath,
and the cutoff frequency of the spectrum of the bath, respectively.
$\theta(\omega)$ is the Heaviside's step function.

Now, we introduce the time-evolution operator $\mathcal{U}(t)$
of the composite system
that obeys the Schr\"{o}dinger equation
\begin{equation}
 i\frac{d}{dt}\mathcal{U}(t)= H(t) \mathcal{U}(t) ,
\end{equation}
and $\mathcal{U}_{\rm S}(t)$ of the system and $\mathcal{U}_{\rm B}(t)$
of the bath as
\begin{equation}
 i\frac{d}{dt}\mathcal{U}_{\rm S}(t) = H_{\rm S}(t)
  \mathcal{U}_{\rm S}(t)
\end{equation}
\begin{equation}
 \mathcal{U}_{\rm B}(t) = e^{-i H_{\rm B} t} \mathcal{U}_{\rm B}(0)
\end{equation}
Using these operators, we define
\begin{equation}
 \mathcal{U}_{\rm int}(t) = \mathcal{U}_{\rm B}^{\dagger}(t)\mathcal{U}_{\rm S}^{\dagger}(t)\mathcal{U}(t) .
\label{eq:Uint_def}
\end{equation}
One can easily see that this obeys
\begin{equation}
 i\frac{d}{dt}\mathcal{U}_{\rm int}(t) = H_{\rm int}^{\rm I}(t) \mathcal{U}_{\rm int}(t) ,
\end{equation}
where $H_{\rm int}^{\rm I}(t)$ is the interaction picture of $H_{\rm int}$
defined by
\begin{equation}
 H_{\rm int}^{\rm I}(t) = \mathcal{U}_{\rm B}^{\dagger}(t)
  \mathcal{U}_{\rm S}^{\dagger}(t)H_{\rm int}\mathcal{U}_{\rm
  S}(t)\mathcal{U}_{\rm B}(t) .
\end{equation}

The density operator is defined by
\begin{equation}
 \rho(t) = \mathcal{U}(t)\rho_{\rm in}\mathcal{U}^{\dagger}(t) ,
\label{eq:rho_def}
\end{equation}
We assume the initial condition as follows.
\begin{equation}
 \rho(0) = \rho_{\rm in} , ~~~
 \rho_{\rm in} = |\Psi_{\rm in}\rangle\langle\Psi_{\rm in}|\otimes
  \frac{e^{-\beta H_{\rm B}}}{Z_{\rm B}} ,
\end{equation}
\begin{equation}
 \mathcal{U}_{\rm S}(0) = 1,~~~ \mathcal{U}_{\rm B}(0) = 1,~~~
  \mathcal{U}_{\rm int}(0) = 1  ,
\end{equation}
where $|\Psi_{\rm in}\rangle$ is the (given) ground state of
$H_{\rm S}(0)$,
$\beta$ is the inverse temperature of the bath,
and $Z_{\rm B}$ is the partition function with respect to the bath,
which is explicitly written as
$Z_{\rm B} = {\rm Tr}_{\rm B}e^{-\beta H_{\rm B}}
= \prod_{j,a}(1 - e^{-\beta\omega_{a}})^{-1}$.
Note that ${\rm Tr}_{\rm B}$ stands for the trace with respect to
the bosonic degree of freedom.

The reduced density operator is defined by
\begin{equation}
 \rho_{\rm S}(t) = {\rm Tr}_{\rm B}\rho(t) .
\end{equation}
Using Eqs.~(\ref{eq:Uint_def}) and (\ref{eq:rho_def}), this writes as
\begin{equation}
 \rho_{\rm S}(t) = \mathcal{U}_{\rm S}(t)\,  {\rm Tr}_{\rm B}
 \left(\mathcal{U}_{\rm int}(t)\rho_{\rm in}
 \mathcal{U}_{\rm int}^{\dagger}(t)\right)
 \mathcal{U}_{\rm S}^{\dagger}(t)
\label{eq:rho_S_1}
\end{equation}

To summarize the setup of the time evolution, 
our composite system is initially in
the product state of the ground state of $H_{\rm S}(0)$
and the thermal equilibrium state of $H_{\rm B}$ at inverse
temperature $\beta$. The spin system and the baths
 starts to interact at $t = 0$, and the composite system
evolves according to
$H(t) = H_{\rm S}(t) + H_{\rm B} + H_{\rm int}$.
The physical quantity of the spin system at $t > 0$ is
determined through the reduced density operator $\rho_{\rm S}(t)$.

\section{Results}\label{sec:Results}

In this section, we present results on 
the kink density
for pure and random Ising chains
computed by our method.
Note that the results on the closed system (namely, $g = 0$)
were obtained by solving the time-dependent
Bogoliubov-de Gennes equation for the equivalent
free fermion model\cite{bib:SuzukiBook}.

The kink density is 
defined by
\begin{equation}
 n := \sum_{j=1}^{N-1}{\rm Tr}_{\rm S}
  \left(\frac{1 - \sigma_j^z\sigma_{j+1}^z}{2}
   \rho_{\rm S}(\tau)\right) ,
\end{equation}
where ${\rm Tr}_{\rm S}$ represents the trace with
respect to the spin degree of freedom,
$\tau$ is the final time, 
$N$ is the number of spins in the system.
This quantity vanishes when the spins are
perfectly aligned along the $z$ axis in the spin space.
Hence the kink density is a measure that quantifies the deviation
from the perfect ferromagnetic state.

We recall here that we assume that in our model the
bosonic operators couple to the longitudinal spin $\sigma_j^z$.
This assumption breaks the integrability of the transverse
Ising chain and has not been considered in the previous studies
for one-dimensional system.
Therefore the results we present below are entirely new.

\subsection{Pure Ising chain}

We first consider quantum annealing of the pure
Ising chain described by the Hamiltonian
\begin{equation}
 H_{\rm S}(t) = - \frac{t}{\tau}\sum_{j=1}^{N-1}
  \sigma_{j}^z\sigma_{j+1}^z
  - \left(1 - \frac{t}{\tau}\right)\sum_{j=1}^N\sigma_j^x ,
\label{eq:H_PureIsing}
\end{equation}
where $\tau$ denotes the runtime of quantum annealing
and the time $t$ evolves from $0$ to $\tau$.
Since this Hamiltonian coincides with
the pure ferromanetic Ising chain at $t = \tau$, the kink
density measures an error of quantum annealing.
We show numerical results on this model with $N = 64$ in
the present subsection.

\begin{figure}[t]
 \includegraphics[width=8cm, bb = 0 0 564 420]{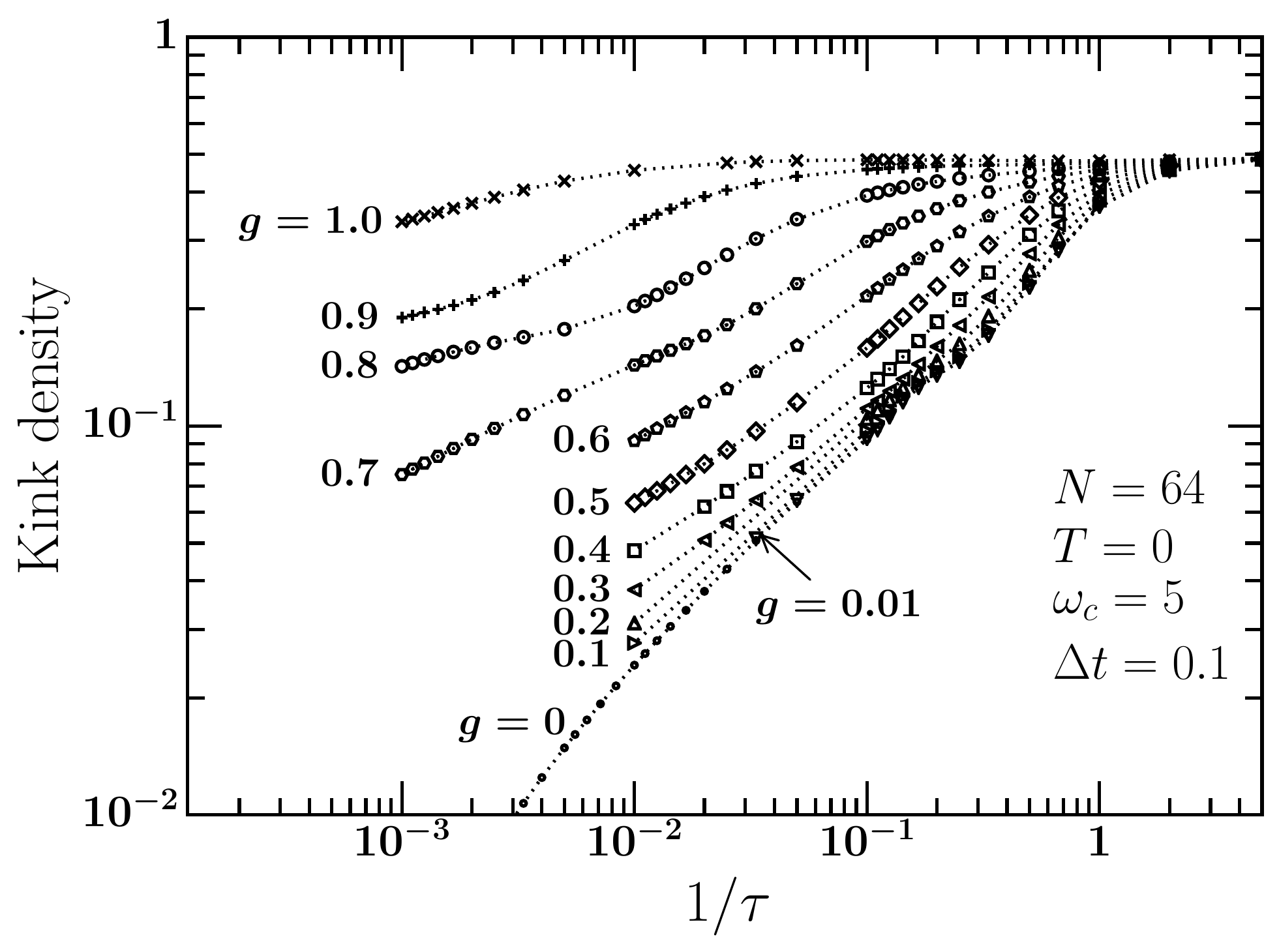}
 \caption{Kink density for the $N=64$ spin system
 with several $g$'s at zero temperature.
 The other parameters are chosen as $\omega_c = 5$, $\mathit{\Delta}t = 0.1$,
 and $l_c = 100$. The numbers of states, $D_t$ and $D_s$,
 kept in the MPS representation are at most $128$.
 The kink density for $g > 0$ is never below
 the value of isolated system (namely, $g = 0$) and grows with
 increasing $g$ at fixed $\tau$.
 The monotonically decaying behavior with increasing $\tau$ is found
 for all $g$'s we studied.}
 \label{fig:N64Eres_T0}
\end{figure}

Let us first consider the dependence of the kink density
on the coupling strength between the system and the bath. 
We present results at $T = 0$ in Fig.~\ref{fig:N64Eres_T0}.
The kink denisty at $T = 0$ increases with inceasing $g$,
implying that coupling to the baths never reduces an
error of quantum annealing compared to the closed system even if the temperature
of the baths is zero. On the other hand, the kink density
decays monotonically with increasing $\tau$ for a fixed $g$.
The slope of this decay becomes smaller for larger $g$.
It is not clear from our results whether
the kink density vanishes for $\tau\to\infty$
when $T = 0$ and $g \neq 0$.


\begin{figure}[t]
 \includegraphics[width=8cm, bb = 0 0 564 420]{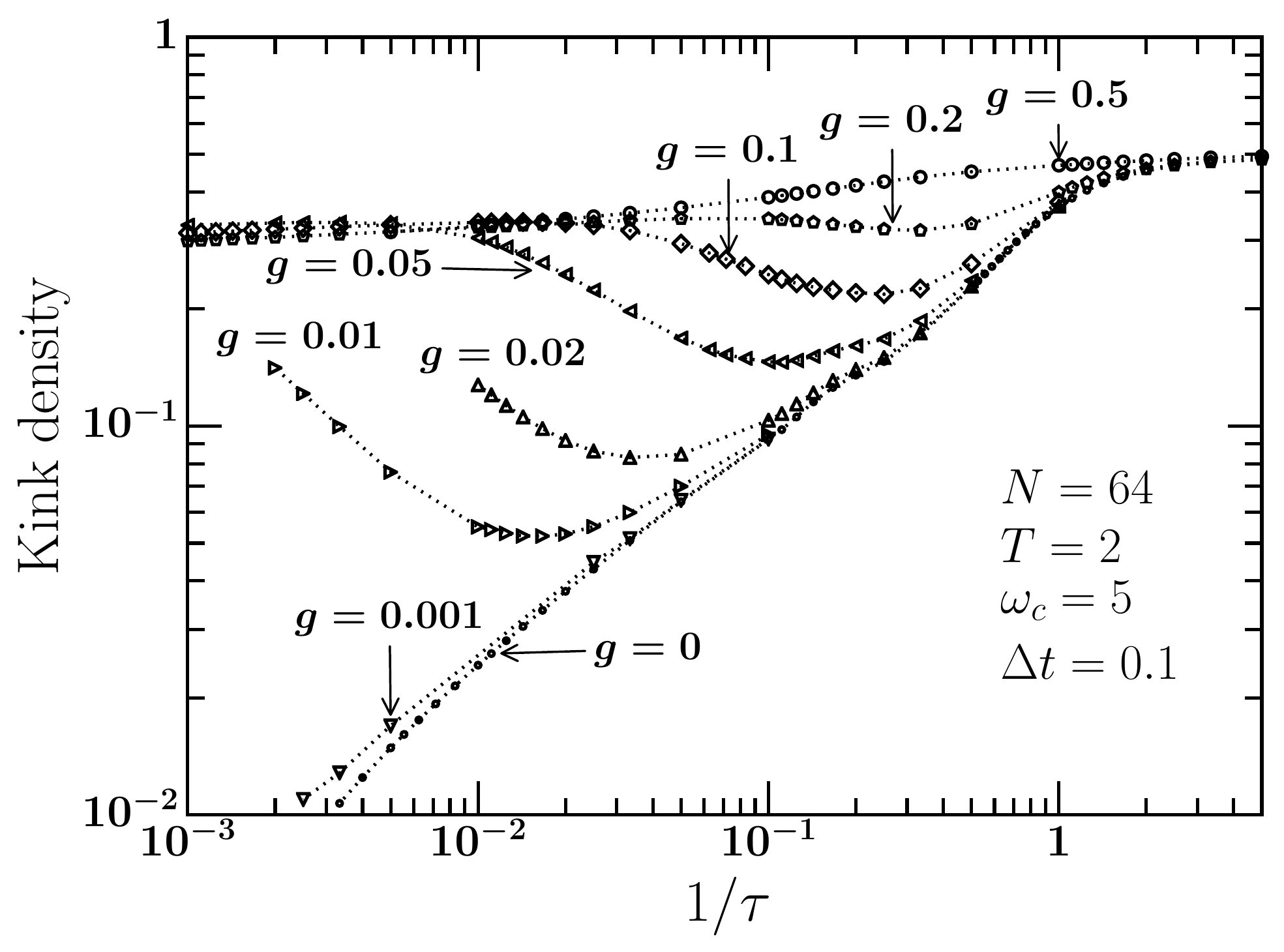}
 \caption{Kink density for the $N=64$ spin system
 with several $g$'s at $T = 2$. The other parameters
 are the same as Fig.~\ref{fig:N64Eres_T0}.
 The kink density grows with $g$, but has a minimum
 as a function of $\tau$. The value of $\tau$
 where the kink density is minimized decreses
 with increasing $g$.
 The results for $g \geq 0.05$ suggest that
 the kink densities for different $g$'s converge with increasing $\tau$.
 }
 \label{fig:N64Eres_T2}
\end{figure}
Figure \ref{fig:N64Eres_T2} shows the dependence of the
kink density on $g$ at $T = 2$. One finds that, at this
temperature,
the kink density turns to an increase with increasing $\tau$
for large $\tau$. We guess that it should be true even for
$g = 0.001$
if one has more data for much larger $\tau$'s.
The minimum position of the kink density
moves to a smaller $\tau$ with increasing $g$. This is because
the coherence time in which the system keeps unaffected by the bath
is shorter for larger $g$. 
Interestingly, the result for $g = 0.2$ suggests that the
kink density decreases again with increasing $\tau$ for very long
$\tau$.
A similar phenomenon has been pointed by Amin in Ref. \citen{bib:AminPRA2015}
and explained by the equilibration with the bath.
We have not yet confirmed whether the equilibration is true or not.
This point is an open issue to be studied.

\begin{figure}[t]
 \includegraphics[width=8cm, bb = 0 0 564 420]{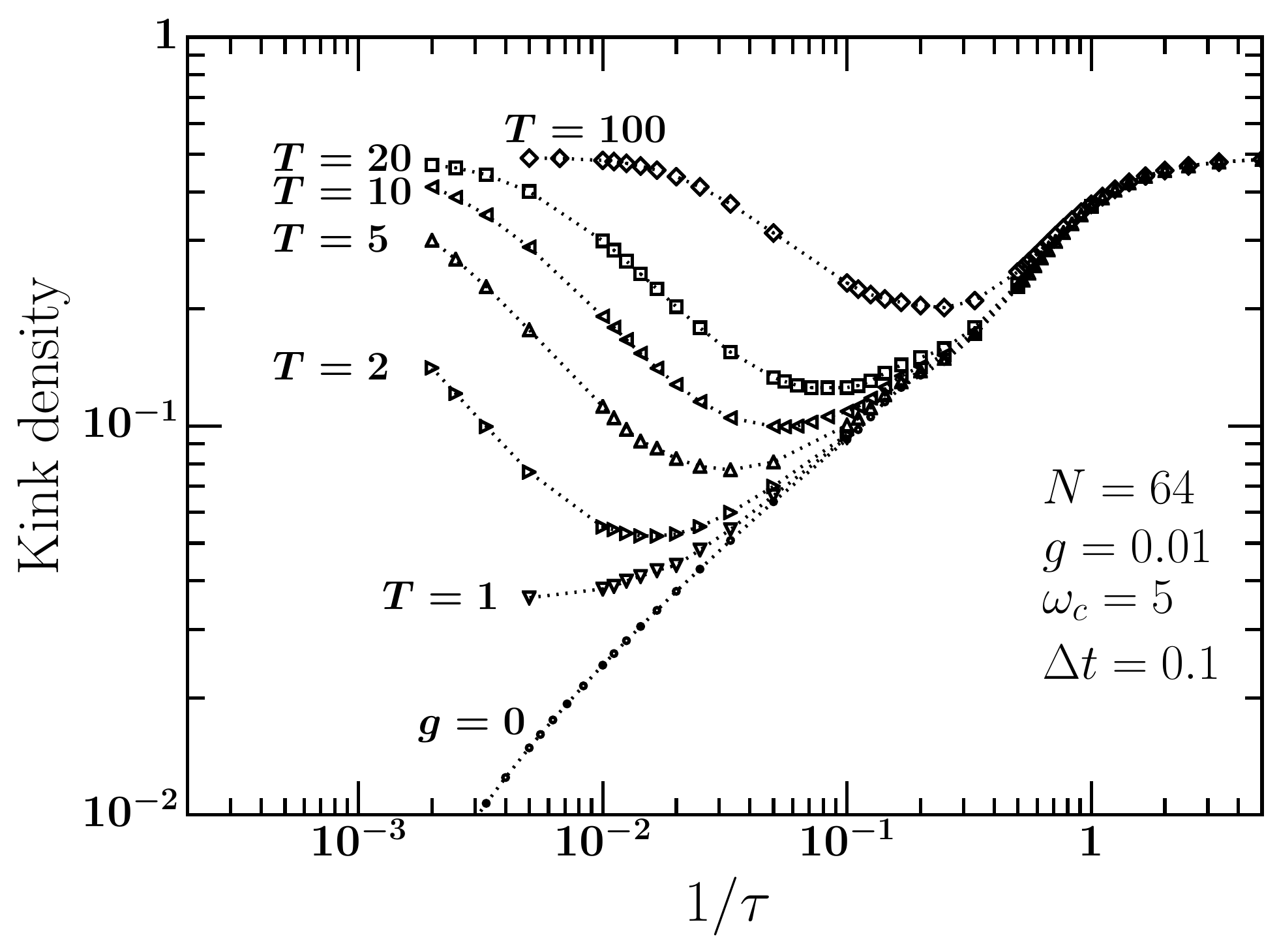}
 \caption{Kink density for the $N = 64$ spin system
 and several temperatures with $g = 0.01$.
 The other parameters
 are the same as Fig.~\ref{fig:N64Eres_T0}.
 The kink density decays with increasing $\tau$ for
 small $\tau$, while it grows with $\tau$ for long $\tau$ at $T \geq 2$.
 At $T = 1$, it is not clear that the kink density grows for long $\tau$.
 The lower the temperature is, the smaller the kink
 density is.
 The kink density is never below the value for the closed system
 ($g = 0$).
 }
 \label{fig:N64Eres_g001}
\end{figure}

Figure \ref{fig:N64Eres_g001} shows results
on the temperature dependence of the kink density for 
the system-bath coupling fixed at $g = 0.01$.
One finds that the kink density starts to grow 
for long $\tau$ at $T \geq 2$, involving a minimum as a function of $\tau$.
This minimum shifts to larger $\tau$ for lower temperature. 
It is not clear from the figure whether there is a minimum at $T = 1$.
Anyway, as shown in Fig.~\ref{fig:N64Eres_T0}, the minimum
disappears when $T = 0$. 
Quite recently, Arceci et al. showed that
the minimum at a certain finite $\tau$ is a global minimum when
$T$ is sufficiently high, but it is in turn a local one with
decreasing the temperature, and disappears finally for sufficiently low $T$
\cite{bib:Arceci2018}.
Although we have not confirmed such transitions, 
there is no conflict between our results and previous ones.

\subsection{Disordered Ising chain}

We here consider quatnum annealing of
the disordered Ising chain described by the Hamiltonian
\begin{equation}
 H_{\rm S}(t) = - \frac{t}{\tau}\sum_{j=1}^{N-1}J_j\sigma_{j}^z\sigma_{j+1}
  - \left(1 - \frac{t}{\tau}\right)\sum_{j=1}^N\sigma_j^x ,
\end{equation}
where the coupling constants $J_j$'s are assumed to be chosen randomly
from the uniform distribution between 0 and 2.
Same as the pure Ising chain, the disordered
Ising chain considered here has the ferromagnetic ground state and
hence the kink density serves to measure an error of quantum annealing.

Although the present disordered Ising chain has a trivial
ground state, excited states or
dynamics are nontrivial and hence this model has drawn a lot of
attentions from the quantum annealing community
as well as statistical physics
community\cite{bib:DziarmagaPRB2006,bib:CanevaPRB2007,bib:SuzukiJSTAT2009,bib:FisherPRB1995,bib:HusePRB2013,bib:KjallPRL2014}.
The excited states of the present system
are characterized by the Anderson localization
when the model is switched to a free fermoin model through
the Jordan-Wigner transformation.
It is a quite interesting problem to inquire into
how the localized state is influenced by the baths.
As for quantum annealing, the localization nature of excited states might change
quantum annealing dynamics from the pure case.
Hence, it is quit important to consider a random model in the
context of quantum annealing.
Here we present a few results on a single instance of this model
with $N = 64$ spins.

\begin{figure}[t]
 \includegraphics[width=8cm, bb = 0 0 842 595]{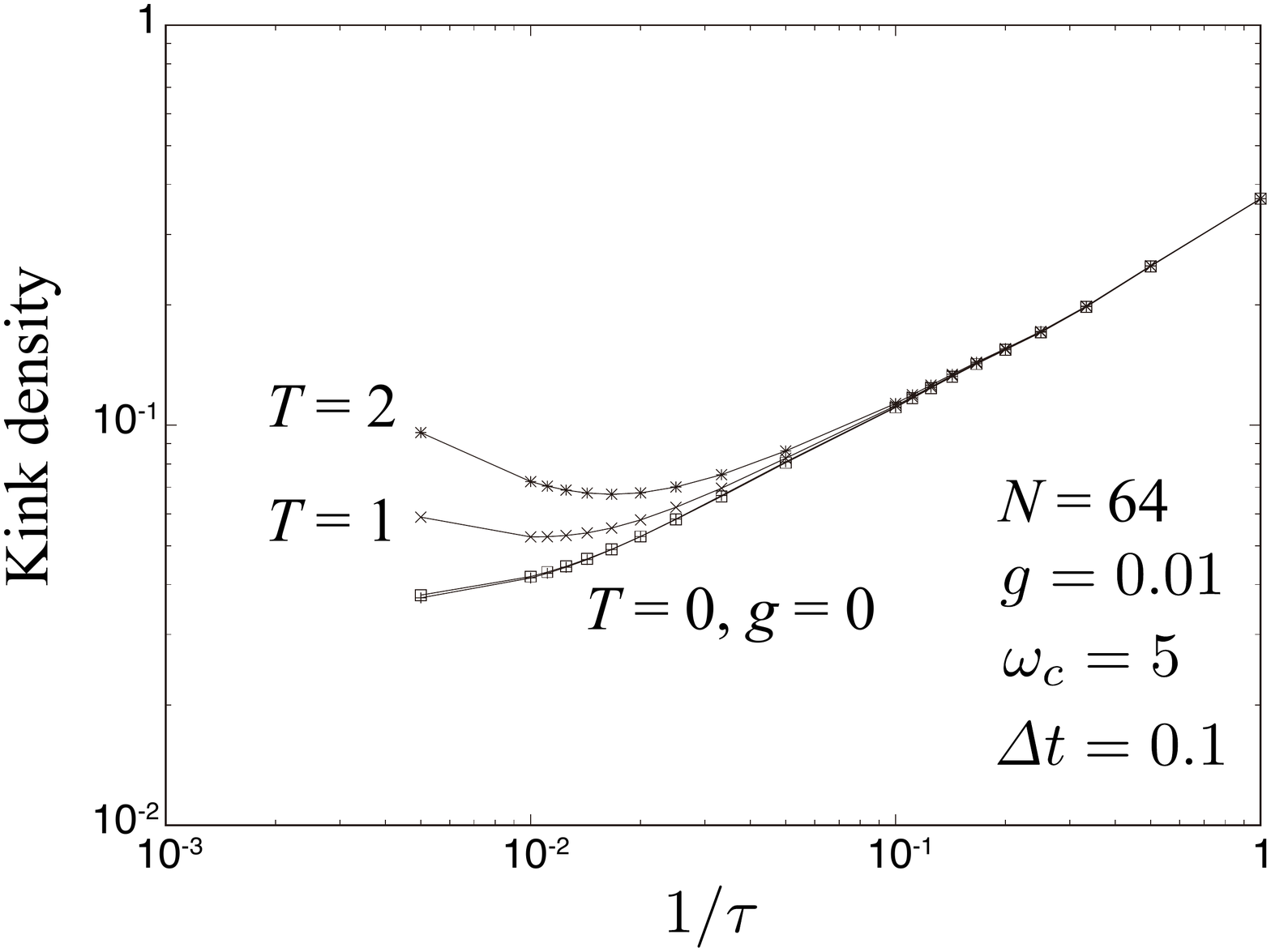}
 \caption{Kink density for a single instance of
 the disorderd Ising chain with $N = 64$ spins.
 The temperature of the bath is set at $T = 0$, $1$, and $2$.
 The other parameters are $g = 0.01$, $\omega_c = 5$,
 $\mathit{\Delta}t = 0.1$, and $l_c = 100$.
 $D_t$ and $D_s$ are up to $64$.
 The kink density at $T = 0$ is very close to 
 that for the closed system. 
 With increasing the temperature, the kink density
 grows and  shows a minimum as a function of $\tau$.
 }
 \label{fig:RandN64Kink_g001}
\end{figure}

Figure \ref{fig:RandN64Kink_g001} shows the results on
the kink density for temperatures $T = 0$, $1$, and $2$.
The system-bath coupling is fixed at $g = 0.01$.
The behavior of the kink density is qualitatively the same
as the pure Ising chain. 
The kink density is never smaller than that for
the closed system even at $T = 0$. With increasing the temperature,
it grows and has a minimum as a function of $\tau$
for $T = 1$ and $2$.
The influence of environment looks qualitatively the same
in pure and random Ising chains on this result.

To summarize the results on the pure and random
Ising chains, coupling to baths never reduces
an error after quantum annealing
from the isolated system,
even if the temperature of the baths is zero.
The negative influence on quantum annealing is more
pronounced with increasing the coupling strength
$g$ between the system and the baths as well as with increasing
the temperature of the baths.
The kink density decays with increasing the annealing runtime
in the limit of low temperature, while it has a minimum when $T$ is
sufficiently high.

\section{Method}\label{sec:Method}

In the present section, we describe the numerical method 
to compute the reduced density matrix of Eq.~(\ref{eq:rho_S_1})
in detail.
The equation of motion of $\rho_{\rm S}(t)$ is too complicated
to solve even for a single spin in general.
When the coupling constant $g^2$ is sufficietly small compared to the
other energy 
scales in the model, the Born-Markov (BM) approximation is
often applied to reduce the difficulty of the problem.
Briefly mentioning, the BM approximation neglects
the order of $g^4$ and higher order terms in the equation
of motion, and replaces the reduced density operator of the past
($\rho_{\rm S}(s)$ with $s < t$) with the current one
($\rho_{\rm S}(t)$)\cite{bib:WeissBook}.
The resulting equation, called as the Redfield equation,
is less complicated than the original one.
However, there are a few problems on this equation or approximation.
(i) The reduced density operator obtained with the BM approximation
has no guarantee that it preserves
the unitarity ${\rm Tr}_S\rho_{\rm S}(t) = 1$, where ${\rm Tr}_{\rm S}$
means the trace with respect to the system's degree of freedom.
(ii) The approximation cannot be improved in the framework
of the BM approximation.
(iii) The Redfield equation reduces to a linear differential
equation for $2^{2N}$ unknown complex functions of $t$.
Therefore the computational cost increases exponentially with the
number $N$ of spins and this method is restricted to small $N$
up to $N\sim 10$.
The first and the third problems might be cured by the
so-called Lindblad approximation. This prescription,
however, makes the second problem more serious.

In the present study,
we develop a novel numerical method in order to avoid the problem
of the BM approximation.
Our approach is based on the path integral and utilizes
the MPS formalism, in which 
the unitarity of the trace is preserved.
It includes a sort of approximation, but the accuracy
of the result
can be improved systematically.
Moreover, systems with $N\sim 10^{2}$ spins are accessible.
One limitation of our method 
is that it is suitable only for
one dimensional system, as is the case with the
DMRG of a closed system.
We describe this method in the following subsections.

\subsection{QUAPI}\label{sec:QUAPI}

We are going to compute the matrix elements of $\rho_{\rm S}(t)$,
starting from Eq.~(\ref{eq:rho_S_1}). At first we focus on the trace with
respect to the bosonic degree of freedom.
This trace can be done by performing the path integral of the
bosonic degree of freedom.
In order to perform the path integral, we apply the
Trotter decomposition to $\mathcal{U}_{\rm int}(t)$.
Letting $M$ be the Trotter number such that $t = M\mathit{\Delta} t$,
the Trotter decomposition yields
\begin{eqnarray} 
 &&\mathcal{U}_{\rm int}(t = M\mathit{\Delta} t) \nonumber\\
 &&= \left[e^{-i H_{\rm int}^{\rm I}(M\mathit{\Delta} t)\frac{\mathit{\Delta} t}{2}}
	  e^{-i H_{\rm int}^{\rm I}((M-1)\mathit{\Delta} t)\frac{\mathit{\Delta}
	  t}{2}}\right] \times \cdots\nonumber\\
 &&  \times
  \left[e^{-i H_{\rm int}^{\rm I}(\mathit{\Delta} t)\frac{\mathit{\Delta} t}{2}}
   e^{-i H_{\rm int}^{\rm I}(0)\frac{\mathit{\Delta} t}{2}}\right]
  + \mathcal{O}(\mathit{\Delta} t^3) \nonumber\\
 &&= e^{-i H_{\rm int}^{\rm I}(M\mathit{\Delta} t)\frac{\mathit{\Delta} t}{2}}
  e^{-i H_{\rm int}^{\rm I}((M-1)\mathit{\Delta} t)\mathit{\Delta} t}
  \cdots
  e^{-i H_{\rm int}^{\rm I}(\mathit{\Delta} t)\mathit{\Delta} t}
  e^{-i H_{\rm int}^{\rm I}(0)\frac{\mathit{\Delta} t}{2}} \nonumber\\
 && + \mathcal{O}(\mathit{\Delta} t^3) .
\label{eq:Trotter}
\end{eqnarray}
Each exponential operator can be written as
\begin{eqnarray}
 &&\hspace{-2em}e^{-iH_{\rm int}^{\rm I}(l\mathit{\Delta} t)\mathit{\Delta}t}\nonumber\\
 &&\hspace{-2em}=
  \mathcal{U}_{\rm S}^{\dagger}(l\mathit{\Delta} t)\left[\prod_{j,a}
   e^{-i\mathit{\Delta} t\sigma_j^z\lambda_{ja}b_{ja}^{\dagger}e^{i\omega_a
   l\mathit{\Delta} t}}
   e^{-i\mathit{\Delta} t\sigma_j^z\lambda_{ja}b_{ja}e^{-i\omega_a l\mathit{\Delta} t}}
   e^{-\frac{1}{2}\mathit{\Delta} t^2\lambda_{ja}^2}
       \right] \nonumber\\
 &&\times \mathcal{U}_{\rm S}(l\mathit{\Delta} t) ,
\label{eq:exponentialOp}
\end{eqnarray}
where $l = 1,\cdots, M-1$. Note an identity
$e^{A + B} = e^A e^B e^{-[A,B]/2}$
for a couple of operators $A$ and $B$ satifying
$[A,B]\in \mathbb{R}$, and the relation
$e^{iH_{\rm B}l\mathit{\Delta} t}b_{ja}e^{-iH_{\rm B}l\mathit{\Delta} t}
= b_{ja}e^{-i\omega_a l\mathit{\Delta} t}$.
Inserting the completeness relation made from the product of an
eigenstate of $\sigma_{j}^z$ and the coherent state of bosons between
the right-sided square bracket and $\mathcal{U}_{\rm S}(t)$ 
in Eq. (\ref{eq:Trotter}),
one obtains the path integral representation for Eq. (\ref{eq:rho_S_1}).
Fortunately, the path integral over the bosonic coherent-state parameter
is Gaussian and can be performed.
See Appendix \ref{sec:App:PI} for details.
The result, after taking the continuous limit of the spectral density,
is given as follows:
\begin{eqnarray}
 &&\rho_{\rm S}(t =
  M\mathit{\Delta}t)\Bigr|_{\sigma_{1,M}\cdots\sigma_{N,M};
  \tau_{1,M}\cdots\tau_{N,M}} \nonumber\\
 &&:= \langle \boldsymbol{\sigma}_{M}|\rho_{\rm S}(M\mathit{\Delta}t)
  |\boldsymbol{\tau}_{M}\rangle \nonumber\\
 &&\cong \mathcal{N}\sum_{\sigma_{1,0}=\pm 1}\sum_{\tau_{1,0}=\pm 1}
  \cdots\sum_{\sigma_{N,M-1}=\pm 1}\sum_{\tau_{N,M-1}=\pm 1}
  \exp\left(\mathcal{H}\right)\mathit{\Psi}_0 ,\nonumber\\
 \label{eq:rho_S_QUAPI}
\end{eqnarray}
where $|\boldsymbol{\sigma}_{M}\rangle$ and
$|\boldsymbol{\tau}_{M}\rangle$ are eigenstates of $\sigma_j^z$ with
eigenvalues $\sigma_{j,M}$ and $\tau_{j,M}$ ($j = 1,2,\cdots,N$),
respectively. The effective
Hamiltonian $\mathcal{H}$ is divided into two parts as
$\mathcal{H} = \mathcal{H}_{\rm S} + \mathcal{H}_{\rm int}$, where
$\mathcal{H}_{\rm S}$ comes from the isolated spin system and $\mathcal{H}_{\rm
int}$ is from the interaction between the spin system and the baths.
They are defined by
\begin{eqnarray}
 &&\mathcal{H}_{\rm S} = \nonumber\\
 && - \frac{i\mathit{\Delta}t}{8}\sum_{l=1}^M
  \left(3 H_l^z(\boldsymbol{\sigma}_l) + 
   H_{l-1}^z(\boldsymbol{\sigma}_l)
  + H_l^z(\boldsymbol{\sigma}_{l-1})+ 3
  H_{l-1}^z(\boldsymbol{\sigma}_{l-1})\right)
  \nonumber\\
 &&+ \frac{i\mathit{\Delta}t}{8}\sum_{l=1}^M
  \left(3 H_l^z(\boldsymbol{\tau}_l) + 
   H_{l-1}^z(\boldsymbol{\tau}_l)
  + H_l^z(\boldsymbol{\tau}_{l-1})+ 3
  H_{l-1}^z(\boldsymbol{\tau}_{l-1})\right)
  \nonumber\\
 && + \sum_{l=1}^M\sum_{j=1}^N\gamma_{l}\sigma_{j,l}\sigma_{j,l-1}
  + \gamma_{l}^{\ast}\tau_{j,l}\tau_{j,l-1}
\label{eq:Heff_S_def}
\end{eqnarray}
\begin{eqnarray}
 &&\mathcal{H}_{\rm int} = \mathit{\Delta}t^2\sum_{j=1}^N\Biggl[\nonumber\\
 && + L\sum_{l=0}^M\sigma_{j,l}\tau_{j,l}
  \nonumber\\
 && - \sum_{M\geq l> m\geq 0}
  \left\{K((l-m)\mathit{\Delta}t)\sigma_{j,l}\sigma_{j,m}+
   K^{\ast}((l-m)\mathit{\Delta}t)\tau_{j,l}\tau_{j,m}\right\}\nonumber\\
 && + \sum_{M\geq l > m\geq 0}
  \left\{K((l-m)\mathit{\Delta}t)\sigma_{j,l}\tau_{j,m}+
   K^{\ast}((l-m)\mathit{\Delta}t)\tau_{j,l}\sigma_{j,m}  \right\}
  \Biggr],\nonumber\\
\label{eq:Heff_int_def}
\end{eqnarray}
respectively. We remark that $\sigma_{j,0}$, $\sigma_{j,M}$, $\tau_{j,0}$, and
$\tau_{j,M}$ with $j = 1,2,\cdots, N$ in $\mathcal{H}_{\rm int}$
must be multiplied by the factor $\frac{1}{2}$.
$H_l^z(\vec{\sigma})$ in Eq. (\ref{eq:Heff_S_def})
denotes the Ising-model part in $H_{\rm S}(t)$ that is written as
\begin{equation}
 H_l^z(\boldsymbol{\sigma}_m) = - \sum_{j=1}^N
  J_j(l\mathit{\Delta}t)\sigma_{j,m}\sigma_{j+1,m} ,
\end{equation}
and $\gamma_{l}$ is defined by
\begin{equation}
 \gamma_{l} = \frac{1}{2}\log\frac{
  \cos [\{ h(l\mathit{\Delta}t) + h((l-1)\mathit{\Delta}t) \}
  \frac{\mathit{\Delta}t}{2}]}
  {i\sin[\{h(l\mathit{\Delta}t) + h((l-1)\mathit{\Delta}t) \}
  \frac{\mathit{\Delta}t}{2}]} .
\label{eq:gamma_l}
\end{equation}
$L$ in Eq. (\ref{eq:Heff_int_def}) is defined by
\begin{equation}
 L = \int_0^{\infty}J(\omega)\frac{1 + e^{-\beta\omega}}{1 -
  e^{-\beta\omega}}d\omega ,
\label{eq:L}
\end{equation}
and $K((l-m)\mathit{\Delta}t)$ is defined by
\begin{equation}
 K((l-m)\mathit{\Delta}t) = \int_0^{\infty}J(\omega)
  \frac{e^{-i\omega(l-m)\mathit{\Delta}t} + e^{-\beta\omega + i\omega(l-m)\mathit{\Delta}t}}
  {1 - e^{-\beta\omega}}d\omega .
\label{eq:Kernel}
\end{equation}
The effecitve wave function $\mathit{\Psi}_0$ comes from the initial spin state,
which is written as
\begin{equation}
 \mathit{\Psi}_0 = \langle\boldsymbol{\sigma}_0|\Psi_{\rm in}\rangle
  \langle\Psi_{\rm in}|\boldsymbol{\tau}_0\rangle .
\label{eq:Psi0_def}
\end{equation}
The normalization factor $\mathcal{N}$ is given by
\begin{eqnarray}
 \mathcal{N} &=& 
  \exp \left[- N\left(M-\frac{1}{2}\right)L\mathit{\Delta}t^2\right]
  \\
  &&\times\prod_{l=1}^M\prod_{j=1}^N
  \left|
  \frac{i}{2}
   \sin \left[ \left\{ h_j(l\mathit{\Delta}t) +
	      h_j((l-1)\mathit{\Delta}t)
	     \right\}\mathit{\Delta}t\right]
       \right| ,\nonumber
\end{eqnarray}
which is necessary to have ${\rm Tr}_{\rm S}\rho_{\rm
S}(M\mathit{\Delta}t) = 1$.

\begin{figure}[t]
 \begin{center}
  \includegraphics[width=8cm, bb=0 0 413 308]{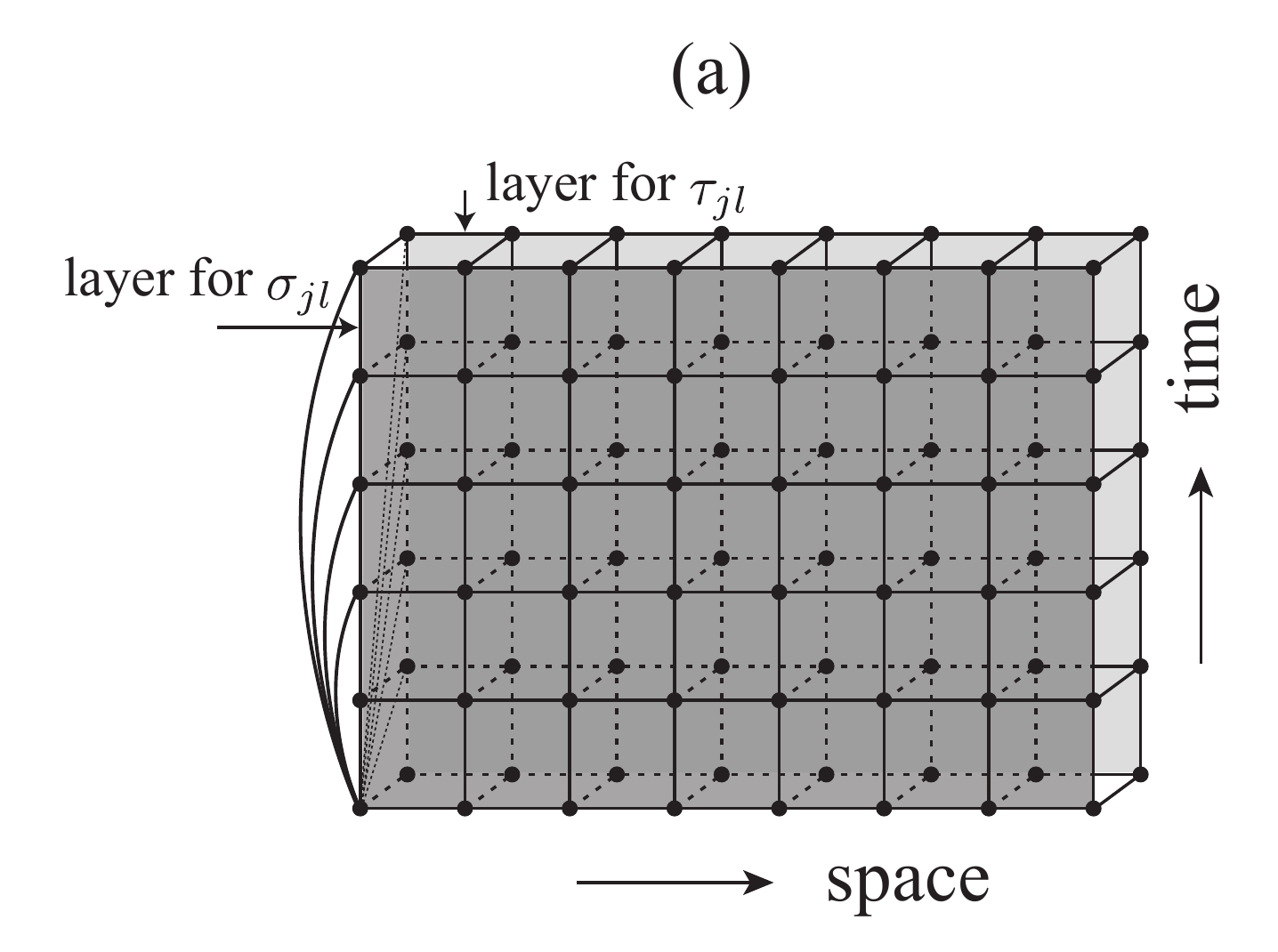}
 \end{center}
 \caption{Lattice and bond structures of a double-layered
 two-dimensional model represented by $\mathcal{H}$. Dots and lines
 represent Ising spins and interactions, respectively.
 Nearest neighbor interactions within each layer comes from
 $\mathcal{H}_{\rm S}$ of a closed system, while interlayar and
 intralayer long-range interactions
 along time direction within the same spatial site are brought from
 the system-bath coupling.
 The interlayer interactions and the long-range interactions
 are not present between different spatial sites.
 Note that only a part of the lines for long-range interactions are
 drawn.
 }
\label{fig:QUAPI_2D}
\end{figure}

One can see from Eqs. (\ref{eq:Heff_S_def}) and
(\ref{eq:Heff_int_def}) that the
effective Hamiltonian $\mathcal{H}$ is equivalent to
a double-layered (1+1)-dimensional Ising model. 
If the system-bath coupling is absent, each layer
is decoupled from the other and the interactions between
different times work only between nearest neighbors within the
same spatial site, see Eq.~(\ref{eq:Heff_S_def}).
In the presence of the system-bath coupling, however,
$\mathcal{H}_{\rm int}$ brings interlayer as well as
intralayer long-range interactions along the time axis
within the same spatial site.
Figure \ref{fig:QUAPI_2D} shows interactions in $\mathcal{H}$
schematically.

We note that the long-range interactions induced by the system-bath
coupling do work within
a spatial site. They never work between spatially sparated spins.
Moreover, the long-range interactions as well as the interlayer
interactions are uniform in space. This is obvious from
Eq.~(\ref{eq:Heff_int_def}) because
their interaction constants do not depend on the site index for space.
These properties are used when we construct a numerical computation
method.

\begin{figure}[t]
\begin{center}
 \includegraphics[width=7cm, bb=0 0 398 232]{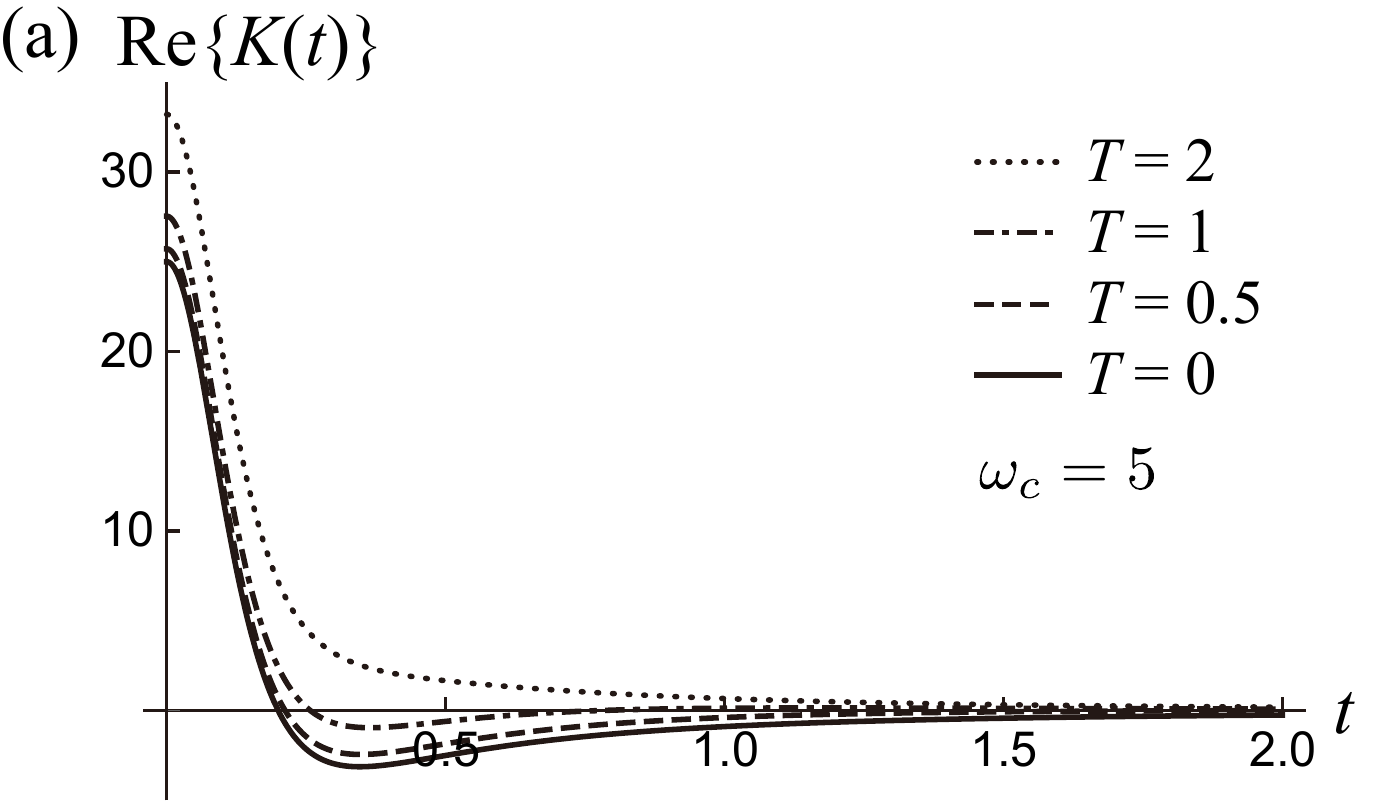}

\vspace{0.5cm}

 \includegraphics[width=7cm, bb=0 0 396 244]{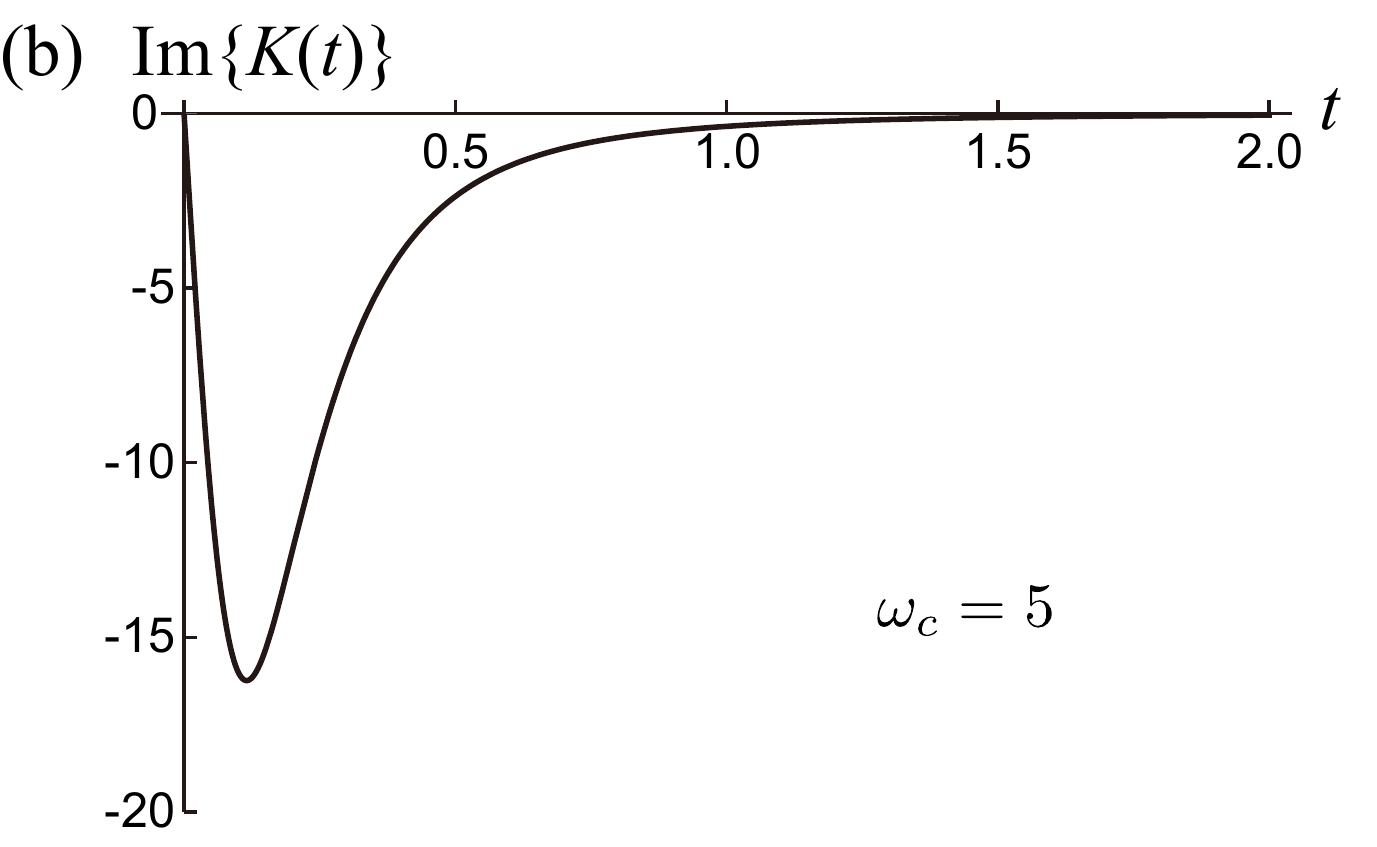}
\end{center}
 \caption{The interaction constant $K(t)$ of the long-range interactions
 along the time direction. (a) The real part of $K(t)$ for
 temperatures, $T = 0$, $0.5$, $1$, and $2$. It turns out
 that the real part decays with $t$.
 One can show that ${\rm Re}\{K(t)\}$ vanishes as $1/t^2$ when $T = 0$
 and as $1/t$ when $T > 0$ for large $t$.
 (b) The imaginary part of $K(t)$. The imaginary part is independent
 of the temperature. It deceases raplidly from zero, takes a minimum and
 vanishes with $t$.
 One can show that ${\rm Im}\{K(t)\}$
 vanishes as $1/t^3$ for large $t$. For both figures,
 we fixed $\omega_c = 5$.
 }
 \label{fig:K}
\end{figure}

The behavior of the long-range interactions given by $K(t)$
along the time direction is shown in Fig. \ref{fig:K}.
The real part decays with $t$ from a large positive value.
It goes down below zero for low temperatures, while it decreases
monotonically towards zero for high temperatures.
For long times, it decays 
as $1/t^2$ for $T = 0$ and $1/t$ for $T > 0$.
The imaginary part, on the other hand, is independent of the
tempratures. It decreases from zero
at first with $t$ and then turns into an increase up to zero.
One can show that the decay for long times is as $1/t^2$.

Now, in order to obtain the matrix elements of the
reduced density matrix, it is necessary to perform the
trace of $\exp({\mathcal{H}})\mathit{\Psi}_0$ with respect to
$\sigma_{j,l}$ and $\tau_{j,l}$ with $j = 1,2,\cdots,N$
and $l = 0,1,\cdots,M-1$. The brute-force method
is useless for this computation, since $e^{NM}$ terms must be
summed up. This computation is reminiscent of
the transfer matrix method in 
the computation of the partition function of the two-dimensional
Ising model or 
that of the time evolution of the transverse Ising model in one
dimension. However, the presence of the long-range interaction
along the time direction makes the problem complicated.
In the next subsection, we introduce a matrix product state representation to
facilitate this computation for large $M$ and $N$.

\subsection{MPS representation}\label{sec:DMRG}

\subsubsection{The time direction}

\begin{figure}[t]
 \begin{center}
  \includegraphics[width=4cm,bb = 4 2 186 225]{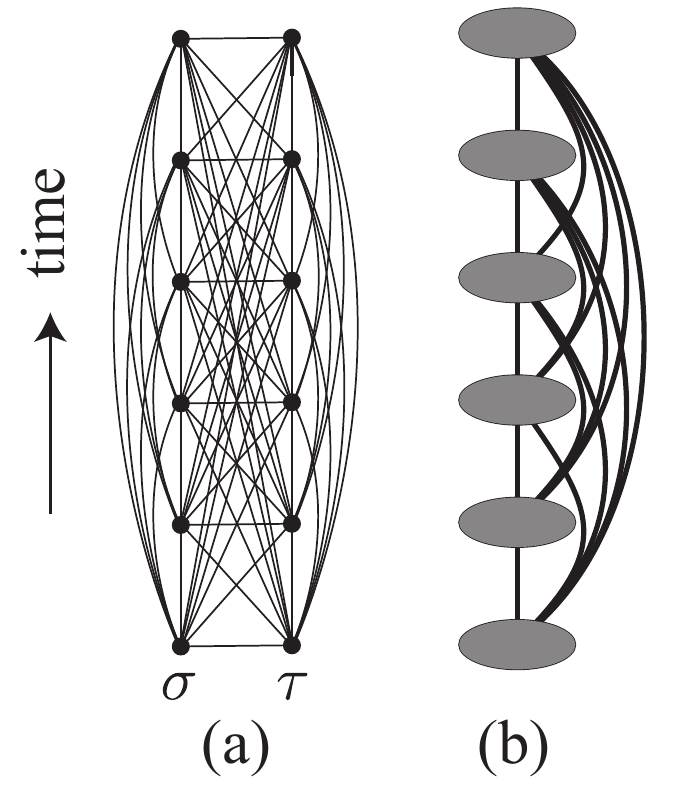}
 \end{center}
 \caption{(a) A ladder of Ising spins along the time direction with a fixed spatial
 site. (b) Chain of pseudospins represented by elliptics. Pseudospin is
 a composite of two Ising spins, $\sigma_{jl}$ and $\tau_{jl}$, with
 the same indices. Interactions due to the systems-bath couping
 and the transverse field are shown by lines in both pictures.
 }
\label{fig:QUAPI_SingleSite}
\end{figure}
Let us focus on a ladder of Ising spins along the time direction
with a fixed spatial site, which consists of $\sigma_{j,l}$
and $\tau_{j,l}$ with a fixed $j$ and $l = 0,1,\cdots M$. 
The interaction originated from the system-bath coupling
works between any pair of Ising spins in this ladder and
is restricted inside the ladder, see Fig.~\ref{fig:QUAPI_SingleSite}(a).
In addition, there are nearest neighbor interactions due to the transverse field.
Now we group $\sigma_{j,l}$ and $\tau_{j,l}$ with the same $l$
(and $j$ as well) and define a pseudospin by this composite.
Remark that this pseudospin has four states.
Then the ladder of Ising spins is seen as a chain of pseudospins
as shown in Fig.~\ref{fig:QUAPI_SingleSite}(b).

Let $S_{j,l}$ denote the state of a pseudospin,
such as $S_{j,l} = (1-\sigma_{j,l})/2 + 2\{(1-\tau_{j,l})/2\}$ for
instance.
Using this variable, the factor of $\exp(\mathcal{H}_{\rm int})$
for the spatial site $j$ is written as
\begin{eqnarray}
 &&\psi_{S_{j,0}S_{j,1}\cdots S_{j,M}} := \exp \Biggl[
  \sum_{l=1}^M\left(\gamma_l\sigma_{j,l}\sigma_{j,l-1}
	       + \gamma^{\ast}_l\tau_{j,l}\tau_{j,l-1}\right)\nonumber\\
 && + \mathit{\Delta}t^2
 \biggl[L\sum_{l=0}^M\sigma_{j,l}\tau_{j,l}
  \nonumber\\
 && - \sum_{M\geq l> m\geq 0}
  \left\{K((l-m)\mathit{\Delta}t)\sigma_{j,l}\sigma_{j,m}+
   K^{\ast}((l-m)\mathit{\Delta}t)\tau_{j,l}\tau_{j,m}\right\}\nonumber\\
 && + \sum_{M\geq l > m\geq 0}
  \left\{K((l-m)\mathit{\Delta}t)\sigma_{j,l}\tau_{j,m}+
   K^{\ast}((l-m)\mathit{\Delta}t)\tau_{j,l}\sigma_{j,m}  \right\}
  \biggr]\Biggr].\nonumber\\
 \label{eq:psi_def}
\end{eqnarray}
This can be seen as a wave function for a one-dimensional
array of $M$ pseudospins with the basis
$S_{j,0},S_{j,1},\cdots,S_{j,M}$, and written in the form of 
a MPS as
\begin{equation}
 \psi_{S_{j,0}S_{j,1}\cdots S_{j,M}}
  = \sum_{\zeta_1,\cdots,\zeta_{M-1}}
  \psi_{S_{j,0}\zeta_1}u^{(1)}_{S_{j,1}\zeta_2;\zeta_1}u^{(2)}_{S_{j,2}\zeta_3;\zeta_2}
  \cdots
  u^{(M-1)}_{S_{j,M-1}S_{j,M};\zeta_{M-1}} ,
  \label{eq:psi_MPS}
\end{equation}
where $u^{(l)}_{S_{j,l}\zeta_{l+1};\zeta_l}$ $(l=1,2,\cdots,M-1)$
and $\psi_{S_{j,0}\zeta_1}$ are defined as follows.

Omitting the site index $j$
the singular value decomposition for $\psi_{S_{0}S_{1}\cdots S_{M}}$
defined the right-sided unitary matrix
$u^{(M-1)}_{S_{M-1}\zeta_{M};\zeta_{M-1}}$: 
\begin{equation}
 \psi_{S_0S_1\cdots S_M}
  = \sum_{\zeta_{M-1}} \phi^{(M-1)}_{S_0S_1\cdots S_{M-2}\zeta_{M-1}}
  \sqrt{\lambda^{(M-1)}_{\zeta_{M-1}}}u^{(M-1)}_{S_{M-1}S_M;\zeta_{M-1}} ,
\label{eq:SVD_t1}
\end{equation}
where $\phi^{(M-1)}_{S_0S_1\cdots S_{M-2}\zeta_{M-1}}$ is a 
unitary matrix and $\sqrt{\lambda^{(M-1)}_{\zeta_{M-1}}}$
is the singular value. 
Similarly, the
singular value decomposition for $\phi^{(l)}$ defines 
$u^{l-1}_{S_{l-1}\zeta_l;\zeta_{l-1}}$ and the singular value 
$\sqrt{\lambda^{(l-1)}_{\zeta_{l-1}}}$ as
\begin{equation}
 \phi^{(l)}_{S_0\cdots S_{l-1}\zeta_{l}}
  = \sum_{\zeta_{l-1}}\phi_{S_0\cdots S_{l-2}\zeta_{l-1}}
  \sqrt{\lambda^{(l-1)}_{\zeta_{l-1}}}u^{(l-1)}_{S_{l-1}\zeta_l;\zeta_{l-1}} ,
\label{eq:SVD_t2}
\end{equation}
with $l=M-1,M-2,\cdots,2$. Finally, we define
$\psi_{S_0\zeta_1}$ as
\begin{equation}
 \psi_{S_0\zeta_1} := \phi^{(1)}_{S_0\zeta_1}\sqrt{\lambda^{(1)}_{\zeta_1}} .
\end{equation}
We assume that the singular values
$\sqrt{\lambda_\zeta^{(l)}}$ ($l=1,2,\cdots$) are ordered as
$\lambda^{(l)}_1 \geq \lambda^{(l)}_2 \geq \cdots$.
As is usual in one-dimensional systems, the singular
value decays rapidly with $l$. In a practical situation,
we introduce a maximum number $D_t$ for the matrix dimension
of $u$ and discard $u^{(l)}_{S_l\zeta_{l+1};l_l}$ with
$l_l > D_t$, so that the error due to tiny singular values
becomes negligible without having the matrix size exponetially
large.
As a result, the matrix product state representation
by Eq.~(\ref{eq:psi_MPS}) reduces the size of $\psi_{S_0S_1\cdots S_M}$
from $4^M$ to $4 D_t^2(M-1)+4 D_t\approx 4 D_t^2M$. This
exponential reduction is the greatest merit of the
MPS representation.

In order to implement the singular value decompositions
in Eqs.~(\ref{eq:SVD_t1}) and (\ref{eq:SVD_t2}), one needs to
take into account all elements of $\psi_{S_0S_1\cdots S_M}$ which
causes a cost exponential in $M$. To avoid this cost, we introduce
the cutoff $l_c$ in the range of the interaction $K(t)$ such that
$K(l\mathit{\Delta}t) = 0$ for $l > l_c$. Since $K(t)$ decays with
$t$, this approximation does not give rise to serious error
if one choose as large $l_c$ as $l_c\mathit{\Delta}t \gg 1$.
Thanks to the interaction cutoff, one can neglect 
the variables $S_m$ with $m < l-l_c$ in the computation of $u^{(l)}$.

\subsubsection{Transfer matrix method}

Having obtained the MPS representation
for a chain of pseudospins along the time direction,
one can perform the trace for the spin degee of freedom.
This computation can be accomplished using the
numerical transfer matrix method combined with the MPS
representation.

We define a factor $B^{[j,m]}_{S_{j,m}S_{j+1,m}}$ for a horizontal bond
in Fig.~\ref{fig:QUAPI_2D} bitween spatial sites $j$ and $j+1$
($j = 1,2,\cdots, N-1$) and temporal site $l$. It is explicitly written as
\begin{eqnarray}
 &&\hspace{-2em}B^{[j,0]}_{S_{j,0}S_{j+1,0}} \\
 &&\hspace{-1em}=
  \exp\left\{
       \frac{i\mathit{\Delta}t}{8}
       \left(J_j(\mathit{\Delta}t) + 3 J_j(0)\right)
       \left(\sigma_{j,0}\sigma_{j+1,0} - \tau_{j,0}\tau_{j+1,0}\right)
      \right\} ,\nonumber\\
 &&\hspace{-2em}B^{[j,l]}_{S_{j,l}S_{j+1,l}} \nonumber\\
 &&\hspace{-1em}=\exp\left\{
       \frac{i\mathit{\Delta}t}{8}
       \left(6J_j(l\mathit{\Delta}t) + J_j((l-1)\mathit{\Delta}t)
	+ J_j((l+1)\mathit{\Delta}t)\right)\right.\nonumber\\
 &&\times\Biggl.
       \left(\sigma_{j,m}\sigma_{j+1,m} - \tau_{j,m}\tau_{j+1,m}\right)
       \Biggr\} ,
       \label{eq:B_m}\\
 &&\hspace{-2em}B^{[j,M]}_{S_{j,M}S_{j+1,M}} \nonumber\\
 &&\hspace{-1em}=\exp\left\{
       \frac{i\mathit{\Delta}t}{8}
       \left(3J_j(M\mathit{\Delta}t) + J_j((M-1)\mathit{\Delta}t)\right)
       \right. \nonumber\\
 &&\times\Biggl.
       \left(\sigma_{j,M}\sigma_{j+1,M} - \tau_{j,M}\tau_{j+1,M}\right)
      \Biggr\} ,
\end{eqnarray}
where $l = 1,2,\cdots, M-1$ in Eq.~(\ref{eq:B_m}).
Using the notation of the quasispins, the reduced density
matrix we are going to compute is expressed as
\begin{eqnarray}
 &&\hspace{-2em}\rho_{\rm S}(M\mathit{\Delta}t)
  \Big|_{\sigma_{1,M}\cdots\sigma_{N,M};\tau_{1,M}\cdots\tau_{N,M}}
  \nonumber\\
 &&\hspace{-2em}= \mathcal{N}
  \prod_{j=1}^{N-1}B^{[j,M]}_{S_{j,M}S_{j+1,M}}
  \sum_{S_{1,M-1}\cdots S_{N,M-1}}
  \prod_{j=1}^{N-1} B^{[j,M-1]}_{S_{j,M-1}S_{j+1,M-1}}
  \nonumber\\
 &&\times\cdots\sum_{S_{1,1}\cdots S_{N,1}}
  \prod_{j=1}^{N-1} B^{[j,1]}_{S_{j,1}S_{j+1,1}}
  \sum_{S_{1,0}\cdots S_{N,0}}
  \prod_{j=1}^{N-1} B^{[j,0]}_{S_{j,0}S_{j+1,0}}
  \nonumber\\
 &&\times
  \prod_{j=1}^N \psi_{S_{j,0}S_{j,1}\cdots S_{j,M}} .
\end{eqnarray}
Note that $\psi_{S_{j,0}S_{j,1}\cdots S_{j,M}}$ is written as
Eq.~(\ref{eq:psi_MPS}).

Let us first consider the trace with respect to
$S_{1,0}$ and $S_{2,0}$. This can be done only by taking
$\psi_{S_{1,0}\zeta_{1,1}}$, $\psi_{S_{2,0}\zeta_{2,1}}$,
$\psi_{S_{3,0}\zeta_{3,1}}$, $B^{[1,0]}_{S_{1,0}S_{2,0}}$,
and $B^{[2,0]}_{S_{2,0}S_{3,0}}$ into account.
We define
\begin{equation}
 \chi^{(2,1)}_{\zeta_{1,1}\zeta_{2,1}S_{3,0}\zeta_{3,1}} :=
  \sum_{S_{1,0},S_{2,0}}
  B^{[1,0]}_{S_{1,0}S_{2,0}}B^{[2,0]}_{S_{2,0}S_{3,0}}
  \psi_{S_{1,0}\zeta_{1,1}}\psi_{S_{2,0}\zeta_{2,1}}\psi_{S_{3,0}\zeta_{3,1}} .
\end{equation}
This can be written in a MPS form as follows.
\begin{equation}
 \chi^{(2,1)}_{\zeta_{1,1}\zeta_{2,1}S_{3,0}\zeta_{3,1}} =
  \sum_{p_{2,1}} v^{(2,1)}_{\zeta_{1,1}\zeta_{2,1};p_{2,1}}
  \sqrt{\kappa^{(2,1)}_{p_{2,1}}}w^{(3,0)}_{S_{3,0}\zeta_{3,1};p_{2,1}} ,
\label{eq:chi21_MPS}
\end{equation}
where $v^{(2,1)}$ and $w^{(3,0)}$ are the left and right-handed unitary matrices
of the singular value decomposition and $\sqrt{\kappa^{(2,1)}_p}$ is the
singular value. Defining
\begin{equation}
 \varphi^{(2,1)}_{p_{2,1}S_{3,0}\zeta_{3,1}}
  := \sum_{\zeta_{1,1},\zeta_{2,1}} (v^{(2,1)}_{\zeta_{1,1}\zeta_{2,1};p_{2,1}})^{\ast}
  \chi^{(2,1)}_{\zeta_{1,1}\zeta_{2,1}S_{3,0}\zeta_{2,1}}
  =
  \sqrt{\kappa^{(2,1)}_{p_{2,1}}}w^{(3,0)}_{S_{3,0}\zeta_{3,1};p_{2,1}}
  , 
\label{eq:varphi21}
\end{equation}
Eq.~(\ref{eq:chi21_MPS}) is arranged into
\begin{equation}
 \chi^{(2,1)}_{\zeta_{1,1}\zeta_{2,1}S_{3,0}\zeta_{3,1}}
  = \sum_{p_{2,1}}v^{(2,1)}_{\zeta_{1,1}\zeta_{2,1};p_{2,1}}
  \varphi^{(2,1)}_{p_{2,1}S_{3,0}\zeta_{3,1}} .
\label{eq:psi_MPS_3}
\end{equation}

Next, we consider the trace with respect to $S_{3,0}$,
multiplying Eq.~(\ref{eq:psi_MPS_3}) by 
$B^{[3,0]}_{S_{3,0}S_{4,0}}$ and $\psi_{S_{4,0}\zeta_{4,1}}$.
Let us define $\chi^{(3,1)}_{p_{2,1}\zeta_{3,1}S_{4,0}\zeta_{4,1}}$ as
\begin{equation}
 \chi^{(3,1)}_{p_{2,1}\zeta_{3,1}S_{4,0}\zeta_{4,1}}
  := \sum_{S_{3,0}}B^{[3,0]}_{S_{3,0}S_{4,0}}
  \varphi^{(2,1)}_{p_{2,1}S_{3,0}\zeta_{3,1}}\psi_{S_{4,0}\zeta_{4,1}} .
\end{equation}
The singular value decomposition leads us to
the following MPS representation:
\begin{equation}
 \chi^{(3,1)}_{p_{2,1}\zeta_{3,1}S_{4,0}\zeta_{4,1}}
  = \sum_{p_{3,1}} v^{(3,1)}_{p_{2,1}\zeta_{3,1};p_{3,1}}
  \varphi^{(3,1)}_{p_{3,1}S_{4,0}\zeta_{4,1}} ,
\end{equation}
where $v^{(3,1)}$ is the left-handed
unitary matrices associated with the singular value decomposition
of $\chi^{(3,1)}$
and $\varphi^{(3,1)}_{p_{3,1}S_{4,0}\zeta_{4,1}}$ is defined in the same manner as
Eq.~(\ref{eq:varphi21}).
Repeating this procedure till $j=N-2$, we obtain
$v^{(j,1)}$ with $j = 2,3,\cdots,N-2$. Finally we carry out
the trace with respect to $S_{N-1,0}$ and $S_{N,0}$, defining
\begin{equation}
 \chi^{(N-1,1)}_{p_{N-2,1}\zeta_{N-1,1}\zeta_{N,1}}
  := \sum_{S_{N-1,0},S_{N,0}}B^{[N-1,0]}_{S_{N-1,0}S_{N,0}}
  \varphi^{(N-2,1)}_{p_{N-2,1}S_{N-1,0}\zeta_{N-1,1}}\psi_{S_{4,0}\zeta_{4,1}} ,
\end{equation}
which in turn writes as 
\begin{equation}
 \chi^{(N-1,1)}_{p_{N-2,1}\zeta_{N-1,1}\zeta_{N,1}}
  = \sum_{p_{N-1,1}} v^{(N-1,1)}_{p_{N-2,1}\zeta_{N-1,1};p_{N-1,1}}
  \varphi^{(N-1,1)}_{p_{N-1,1}\zeta_{N,1}} .
\end{equation}
Thus we obtain
\begin{eqnarray}
 &&\hspace{-2em}\sum_{S_{1,0}\cdots S_{N,0}}\prod_{j=1}^{N-1}
  B^{[j,0]}_{S_{j,0}S_{j+1,0}}\prod_{j=1}^N\psi_{S_{j,0}S_{j,1}\cdots S_{j,M}}
  \nonumber\\
 &&\hspace{-1em}=\sum_{p_{2,1},p_{3,1},\cdots,p_{N-1,1}}
  v^{(2,1)}_{\zeta_{1,1}\zeta_{2,1};p_{2,1}}v^{(3,1)}_{p_{2,1}\zeta_{3,1};p_{3,1}}
  \cdots
  v^{(N-1,1)}_{p_{N-2,1}\zeta_{N-1,1};p_{N-1,1}}
  \varphi^{(N-1,1)}_{p_{N-1,1}\zeta_{N,1}}
  \nonumber\\
 &&\hspace{-1em}=: \Lambda^{(1)}_{\zeta_{1,1}\zeta_{2,1}\cdots\zeta_{N,1}}
  .
  \label{eq:MPS_0_Nb}
\end{eqnarray}
By means of this procedure, one can easily obtain another
MPS representation as follows:
\begin{eqnarray}
 &&\hspace{-4em}\Lambda^{(1)}_{\zeta_{1,1}\zeta_{2,1}\cdots \zeta_{N,1}}
 = \sum_{q_{2,1},q_{3,1}\cdots,q_{N-1,1}}
 \tilde{\varphi}^{(1,1)}_{\zeta_{1,1}q_{2,1}} \nonumber\\
 &&\hspace{2em}\times
  \tilde{w}^{(2,1)}_{\zeta_{2,1}q_{3,1};q_{2,1}} \cdots
  \tilde{w}^{(N-2,1)}_{\zeta_{N-2,1}q_{N-1,1};q_{N-2,1}}
  \tilde{w}^{(N-1,1)}_{\zeta_{N-1,1}l_{N,1};q_{N-1,1}} ,
  \label{eq:MPS_0_Nf}
\end{eqnarray}
where $\tilde{w}^{(j,1)}$'s and
$\tilde{\varphi}^{(1,1)}_{\zeta_{1,1}q_{2,1}}$
are the right-handed unitary matrices associated with
the singular value decompositions.
\begin{eqnarray}
 &&\hspace{-2em}
  \sum_{p_{N-1,1}}v^{(N-1,1)}_{p_{N-2,1}\zeta_{N-1,1};p_{N-1,1}}
  \varphi^{(N-1,1)}_{p_{N-1,1}\zeta_{N,1}}
  \nonumber\\
 &&= \sum_{q_{N-1,1}}\tilde{v}^{(N-2,1)}_{p_{N-2,1};q_{N-1,1}}
  \sqrt{\tilde{\kappa}^{(N-1,1)}_{q_{N-1,1}}}
  \tilde{w}^{(N-1)}_{\zeta_{N-1,1}\zeta_{N,1};q_{N-1,1}} \nonumber\\
 &&=: \sum_{q_{N-1,1}}\tilde{\varphi}^{(N-1,1)}_{p_{N-2,1}q_{N-1,1}}
  \tilde{w}^{(N-1,1)}_{\zeta_{N-1,1}\zeta_{N,1};q_{N-1,1}} ,
\end{eqnarray}
and
\begin{eqnarray}
  \sum_{p_{j,1}}v^{(j,1)}_{p_{j-1,1}\zeta_{j,1};p_{j,1}}\tilde{\varphi}^{(j,1)}_{p_{j,1}q_{j+1,1}}
 &=& \sum_{q_{j,1}}\tilde{v}^{(j-1,1)}_{p_{j-1,1};q_{j,1}}
  \sqrt{\tilde{\kappa}^{(j,1)}_{q_j,1}}\tilde{w}^{(j,1)}_{\zeta_{j,1}q_{j+1,1};q_{j,1}}
  \nonumber\\
 &=:& \sum_{q_{j,1}}
  \tilde{\varphi}^{(j-1,1)}_{p_{j-1,1}q_{j,1}}\tilde{w}^{(j,1)}_{\zeta_{j,1}q_{j+1,1};q_{j,1}} 
\end{eqnarray}
for $j = N-2,N-3,\cdots,2$.

Now we move on to the trace on $S_{j,1}$ with $j = 1,2,\cdots N$.
To this end,
one must consider $u^{(1)}_{S_{1,1}\zeta_{1,2};\zeta_{1,1}}
\cdots u^{(1)}_{S_{N,1}\zeta_{N,2};\zeta_{N,1}}$. Suppose
that $\lambda^{(2)}_{\zeta_{1,2}\cdots\zeta_{N,2}}$
is written as
\begin{eqnarray}
 \Lambda^{(2)}_{\zeta_{1,2}\cdots \zeta_{N,2}}
  &:=& \sum_{S_{1,1},\cdots,S_{N,1}}\prod_{j=1}^{N-1}
  B^{[j,1]}_{S_{j,1}S_{j+1,1}}
  \sum_{\zeta_{1,1}\cdots \zeta_{N,1}}
  \Lambda^{(1)}_{\zeta_{1,1}\cdots\zeta_{N,1}}\nonumber\\
 &&\times u^{(1)}_{S_{1,1}\zeta_{1,2};\zeta_{1,1}}
  \cdots u^{(1)}_{S_{N,1}\zeta_{N,2};\zeta_{N,1}}  .
\label{eq:MPS_1}
\end{eqnarray}
This factor can be written in the MPS form
by performing the traces on $S_{j,1}$ with $j = 1,2,\cdots, N$
and the singular value decompositions recursively. 
\begin{eqnarray}
 \Lambda^{(2)}_{\zeta_{1,2}\cdots \zeta_{N,2}}
 &=& \sum_{p_{2,2},p_{3,2}\cdots,p_{N-1,2}}
 v^{(2,2)}_{\zeta_{1,2}\zeta_{2,2};p_{2,2}}
 v^{(3,2)}_{\zeta_{2,2}\zeta_{3,2};p_{3,2}}\cdots
 v^{(N-1,2)}_{p_{N-2,2}\zeta_{N-1,2};p_{N-1,2}}\nonumber\\
 &&\hspace{5em}\times \tilde{\varphi}^{(N-1,2)}_{p_{N-1,2}l_{N,2}}  ,
\label{eq:MPS_1_Nf}
\end{eqnarray}
where $v^{(j,2)}$ and $\varphi^{(j,2)}$with $j = 2,\cdots,N-1$ are defined
through the singular value decompositions as follows: 
\begin{eqnarray}
 &&\hspace{-2em}\sum_{S_{1,1},S_{2,1}}
  \sum_{\zeta_{1,1},\zeta_{2,1},\zeta_{3,1}}\sum_{q_{2,1},q_{3,1}}
  \tilde{\varphi}^{(1,1)}_{\zeta_{1,1}q_{2,1}}
  \tilde{w}^{(2,1)}_{\zeta_{2,1}q_{3,1};q_{2,1}}
  \tilde{w}^{(3,1)}_{\zeta_{3,1}q_{4,1};q_{3,1}}
  B^{[1,1]}_{S_{1,1}S_{2,1}}B^{[2,1]}_{S_{2,1}S_{3,1}}  \nonumber\\
 &&\hspace{-2em}\times u^{(1)}_{{S_{1,1}}\zeta_{1,2};\zeta_{1,1}}
  u^{(1)}_{S_{2,1}\zeta_{2,2};\zeta_{2,1}}
  u^{(1)}_{S_{3,1}\zeta_{3,2};\zeta_{3,1}}
  \nonumber\\
 &&= \sum_{p_{2,2}}v^{(2,2)}_{\zeta_{1,2}\zeta_{2,2};p_{2,2}}
  \sqrt{\kappa^{(2,2)}_{p_{2,2}}}
  w^{(3,1)}_{S_{3,1}\zeta_{3,2}q_{4,1};p_{2,2}} \nonumber\\
 &&=: \sum_{p_{2,2}}v^{(2,2)}_{\zeta_{1,2}\zeta_{2,2};p_{2,2}}
  \varphi^{(2,2)}_{p_{2,2}S_{3,1}\zeta_{3,2}q_{4,1}} ,
\end{eqnarray}
\begin{eqnarray}
 &&\hspace{-2em}\sum_{S_{j,1}}\sum_{l_{j+1,1}}\sum_{q_{j+1,1}}
  \varphi_{p_{j-1,1}S_{j,1}\zeta_{j,2}q_{j+1,1}}
  \tilde{w}^{(j+1,1)}_{\zeta_{j+1,1}q_{j+2,1};q_{j+1,1}}
  B^{[j,1]}_{S_{j,1}S_{j+1,1}}
  u^{(1)}_{{S_{j+1,1}}\zeta_{j+1,2};\zeta_{j+1,1}}
  \nonumber\\
 &&\hspace{-1em} = \sum_{p_{j,2}}v^{(j,2)}_{p_{j-1,2}\zeta_{j,2};p_{j,2}}
  \sqrt{\kappa^{(j,2)}_{p_{j,2}}}
  w^{(j+1,1)}_{S_{j+1,1}\zeta_{j+1,2}q_{j+2,1};p_{j,2}} \nonumber\\
 &&\hspace{-1em} =: \sum_{p_{j,2}}v^{(j,2)}_{p_{j-1,2}\zeta_{j,2};p_{j,2}}
  \varphi^{(j,2)}_{p_{j,2}S_{j+1,1}\zeta_{j+1,2}q_{j+2,1}} ~~~
  (j = 3,\cdots, N-2),
\end{eqnarray}
\begin{eqnarray}
 &&\hspace{-7em}\sum_{S_{N-1,1}S_{N,1}}\sum_{\zeta_{N,1}}
  \varphi^{(N-2,2)}_{p_{N-2,1}S_{N-1,1}\zeta_{N-1,2}\zeta_{N,1}}
  B^{[N-1,1]}_{S_{N-1,1}S_{N,1}} 
  u^{(1)}_{{S_{N,1}}\zeta_{N,2};\zeta_{N,1}} \nonumber\\
 &&\hspace{-5em}= \sum_{p_{N-1,2}}v^{(N-1,2)}_{p_{N-2,2}\zeta_{N-1,2};p_{N-1,2}}
  \sqrt{\kappa^{(N-1,2)}_{p_{N-1,2}}}
  w^{(N,1)}_{\zeta_{N,2};p_{N-1,2}} 
  \nonumber\\
 &&\hspace{-5em}=: \sum_{p_{N-1,2}}v^{(N-1,2)}_{p_{N-2,2}\zeta_{N-1,2};p_{N-1,2}}
  \varphi^{(N-1,2)}_{p_{N-1,2}\zeta_{N,2}} .
\end{eqnarray}
Equation~(\ref{eq:MPS_1_Nf}) is arranged into another
MPS form in a similar manner to obtain
Eq.~(\ref{eq:MPS_0_Nf}) from Eq.~(\ref{eq:MPS_0_Nb}) as
\begin{eqnarray}
 &&\hspace{-4em}\Lambda^{(2)}_{\zeta_{1,2}\cdots \zeta_{N,2}}
 = \sum_{q_{2,2},q_{3,2}\cdots,q_{N-1,2}}
 \tilde{\varphi}^{(1,2)}_{\zeta_{1,2}q_{2,2}} \\
\label{eq:MPS_1_Nb}
 &&\hspace{1em}\times \tilde{w}^{(2,2)}_{\zeta_{2,2}q_{3,2};q_{2,2}}\cdots
  \tilde{w}^{(N-2,2)}_{\zeta_{N-2,2}q_{N-1,2};q_{N-2,2}}
  \tilde{w}^{(N-1,2)}_{\zeta_{N-1,2}\zeta_{N,2};q_{N-1,2}} .\nonumber
\end{eqnarray}

Repeating the same procedure from Eq.~(\ref{eq:MPS_1}) to
(\ref{eq:MPS_1_Nb}), one can accomplish the trace with
respect to $S_{jl}$ with $j = 1,2,\cdots, N$
and $l = 2,\cdots, M-1$ to have
\begin{eqnarray}
 &&\hspace{-3em}\Lambda^{(M)}_{S_{1,M}\cdots S_{N,M}}
 = \sum_{q_{2,M},\cdots,q_{N-1,M}}
  \tilde{\varphi}^{(1,M)}_{S_{1,M}q_{2,M}}
  \\
 &&\hspace{2em}\times
  \tilde{w}^{(2,M)}_{S_{2,M}q_{3,M};q_{2,M}}\cdots
  \tilde{w}^{(N-2,M)}_{S_{N-2,M}q_{N-1,M};q_{N-2,M}}
  \tilde{w}^{(N-1,M)}_{S_{N-1,M}S_{N,M};q_{N-1,M}} .\nonumber
\end{eqnarray}
We finally take into account $B^{[j,M]}_{S_{j,M}S_{j+1,M}}$ 
and obtain
\begin{eqnarray}
 &&\hspace{-2em}\Lambda_{S_{1,M}\cdots S_{N,M}}
 := \prod_{j=1}^{N-1}B^{[j,M]}_{S_{j,M}S_{j+1,M}}
 \Lambda^{(M)}_{S_{1,M}\cdots S_{N,M}}\nonumber\\
 &&= \sum_{p_2,p_3,\cdots,p_{N-1}}
  v^{(2)}_{S_{1,M}S_{2,M};p_{2}}v^{(3)}_{p_{2}S_{3,M};p_{3}}
  \cdots v^{(N-1)}_{p_{N-2}S_{N-1,M};p_{N-1}}\varphi^{(N-1)}_{p_{N-1}S_{N,M}}
  ,\nonumber\\
 &&
\label{eq:MPS_final}
\end{eqnarray}
where $v^{(j)}$ and $\varphi^{(N)}$ are defined through the
singular value decompositions as follows:
\begin{eqnarray}
 &&\hspace{-2em}\sum_{q_{2,M}q_{3,M}}B^{[1,M]}_{S_{1,M}S_{2,M}}B^{[2,M]}_{S_{2,M}S_{3,M}}
  \tilde{\varphi}_{S_{1,M}q_{2,M}}
  \tilde{w}^{(2,M)}_{S_{2,M}q_{3,M};q_{2,M}}
  \tilde{w}^{(3,M)}_{S_{3,M}q_{4,M};q_{3,M}}\nonumber\\
 &&= \sum_{p_2}v^{(2)}_{S_{1,M}S_{2,M};p_{2}}\sqrt{\kappa_{p_2}}
  w^{(3)}_{S_{3,M}q_{4,M};p_{2}} \nonumber\\
 &&=: \sum_{p_2}v^{(2)}_{S_{1,M}S_{2,M};p_{2}}
  \varphi^{(2)}_{p_2S_{3,M}q_{4,M}} ,\\
 &&\hspace{-2em}\sum_{q_{j+1,M}}B^{[j,M]}_{S_{j,M}S_{j+1,M}}
  \varphi^{(j-1)}_{p_{j-1}S_{j,M}q_{j+1,M}}
  \tilde{w}^{(j+1,M)}_{S_{j+1,M}q_{j+2,M};q_{j+1,M}}
  \nonumber\\
 &&= \sum_{p_j}v^{(j)}_{p_{j-1}S_{j,M};p_{j}}\sqrt{\kappa_{p_j}}
  w^{(j+1)}_{S_{j+1,M}q_{j+2,M};p_j} \nonumber\\
 &&=: \sum_{p_j}v^{(j)}_{p_{j-1}S_{j,M};p_{j}}
  \varphi^{(j)}_{p_jS_{j+1,M}q_{j+2,M}} ~~~~(j=3,\cdots, N-2).
  \nonumber\\
 &&
\end{eqnarray}
Equation (\ref{eq:MPS_final}) is the very MPS formula
that we want.

Using Eq.~(\ref{eq:MPS_final}), one can compute several
quantities. Assuming
$S_{j,M} = (1-\sigma_{j,M})/2 + 2\{(1-\tau_{j,M})/2\}$,
the trace of the reduced density matrix is written as
\begin{eqnarray}
 {\rm Tr}_{\rm S}\left(\rho_{\rm S}(t =
  M\mathit{\Delta}t)\right) 
 &=& \sum_{\sigma_{1,M},\cdots,\sigma_{N,M}}
 \rho_{\rm S}(M\mathit{\Delta}t)
 \Bigr|_{\sigma_{1,M}\cdots\sigma_{N,M};
 \sigma_{1,M}\cdots\sigma_{N,M}} \nonumber \\
 &=& \mathcal{N}\sum_{S_{1,M}=0,3}\cdots\sum_{S_{N,M}=0,3}
  \Lambda_{S_{1,M}\cdots S_{N,M}} .
\end{eqnarray}
The energy expectation value at $t = M\mathit{\Delta} t$ is
given by
\begin{eqnarray}
 &&\hspace{-2em}{\rm Tr}_{\rm S}\left(H_{\rm S}
		  \rho_{\rm S}(M\mathit{\Delta}t)\right)\\
 &&= - \sum_{j=1}^{N-1} J_j(M\mathit{\Delta}t)
 {\rm Tr}_{\rm S}\left(\sigma_j^z\sigma_{j+1}^z
		  \rho_{\rm S}(M\mathit{\Delta}t)\right)
 \nonumber\\
 &&\hspace{1em}- h(M\mathit{\Delta}t)\sum_{j=1}^N{\rm Tr}_{\rm S}
  \left(\sigma_j^x \rho_{\rm S}(M\mathit{\Delta}t)\right)
 \nonumber\\
 &&=  - \sum_{j=1}^{N-1} J_j(M\mathit{\Delta}t)
  \sum_{S_{1,M}=0,3}\cdots\sum_{S_{N,M}=0,3}(-1)^{S_{j,M}+S_{j+1,M}}
  \Lambda_{S_{1,M}\cdots S_{N,M}}\nonumber\\
 &&\hspace{1em} - h(M\mathit{\Delta}t)\sum_{j=1}^N
  \sum_{S_{1,M}=0,3}\cdots\sum_{S_{j,M}=1,2}\cdots
  \sum_{S_{N,M}=0,3}
  \Lambda_{S_{1,M}\cdots S_{N,M}} .\nonumber\\
 &&
\label{eq:E-formula}
\end{eqnarray}
Finally, the ground-state probability that the ground state
of the system at time $t=M\mathit{\Delta}t$
is found in the state after time evolution is given by
\begin{equation}
 P_{\rm G} = {\rm Tr}_{\rm S}
  \left(
   |\Psi_{\rm G}\rangle\langle\Psi_{\rm G}
   \rho_{\rm S}(M\mathit{\Delta}t)  \right)
  = \langle\Psi_{\rm G}|
  \rho_{\rm S}(M\mathit{\Delta}t)|\Psi_{\rm G}\rangle ,
\label{eq:P_G-formula}
\end{equation}
where $|\Psi_{\rm G}\rangle$ denotes the ground state
of $H_{\rm S}(t = M\mathit{\Delta}t)$. Note that, if there are
degenerated ground states at $t=M\mathit{\Delta}t$,
it is necessary to add the above quantities computed for
each ground state. For instance, when the ground states
are the fully polarized states along the $\sigma^z$ axis,
the ground-state probability is written simply as
\begin{equation}
 P_{\rm G} = \Lambda_{00\cdots 0} + \Lambda_{33\cdots 3} .
\end{equation}

We comment on the matrix dimension of $v^{(jl)}$,
$\tilde{v}^{(jl)}$, $w^{(jl)}$ and $\tilde{w}^{(jl)}$,
or the range of indices $p_{jl}$ and $q_{jl}$ in other words. 
Speaking regorously, the maximum of matrix size grows exponentially
with the number of spin $N$. However, 
in a practical situation, 
the matrix dimension is restricted to a number $D_{\rm s}$, 
omitting the bases corresponding vanishingly small singular values. 
The faster the singular value decays with increasing
its index, the smaller $D_{\rm s}$ can be. Note that
the singular values are ordered in a descending way.
Therefore the decaying behavior of the singular values
is crucial to the present method.
In fact, it has been well known that, due to the
area law of the entanglement entropy, 
$D_{\rm s}$ can be made so small
in the computation of the ground state in one dimensional system.
As for our problem of a time-dependent open system in one dimension,
although there is no theoretical ground by now, we
expect that an acceptable $D_{\rm s}$ to the numerical computation
suffices to obtain an accurate result.

\subsection{Test on a single spin}

Let us first look through a single spin for the sake of
a test of our method. We consider the Landau-Zener
model coupled to a bosonic bath. The Hamiltonian for
the spin system is given by
\begin{equation}
 H_{\rm LZ}(t) = - \frac{vt}{2}\sigma^z - \frac{1}{2}\sigma^x ,
\end{equation}
where $v$ denotes the velocity of driving.
The Hamiltonians for the bath and the system-bath coupling
are given by Eq.~(\ref{eq:H_B}) and (\ref{eq:Hint}) with omission
of $j$.

The well-known solution of the Landau-Zener model without the bath is
described briefly as follows. Let $|\uparrow\rangle$ and $|\downarrow\rangle$ be
the eigenstates of $\sigma^z$ with the eigenvalues $+1$ and $-1$,
respectively. Assuming the initial condition
$|\psi(t = -\infty)\rangle = |\downarrow\rangle$
which is the ground state of $H_{\rm LZ}(t = -\infty)$,
the ground state probability denoted by $P_{\rm G}$
that the system remains in the ground state at $t = +\infty$
is given by
$P_{\rm G}^{\rm LZ} = 1 - \exp(-\pi/2v)$.

\begin{figure}[t]
 \begin{center}
  \includegraphics[width=8cm, bb = 0 1 564 420]{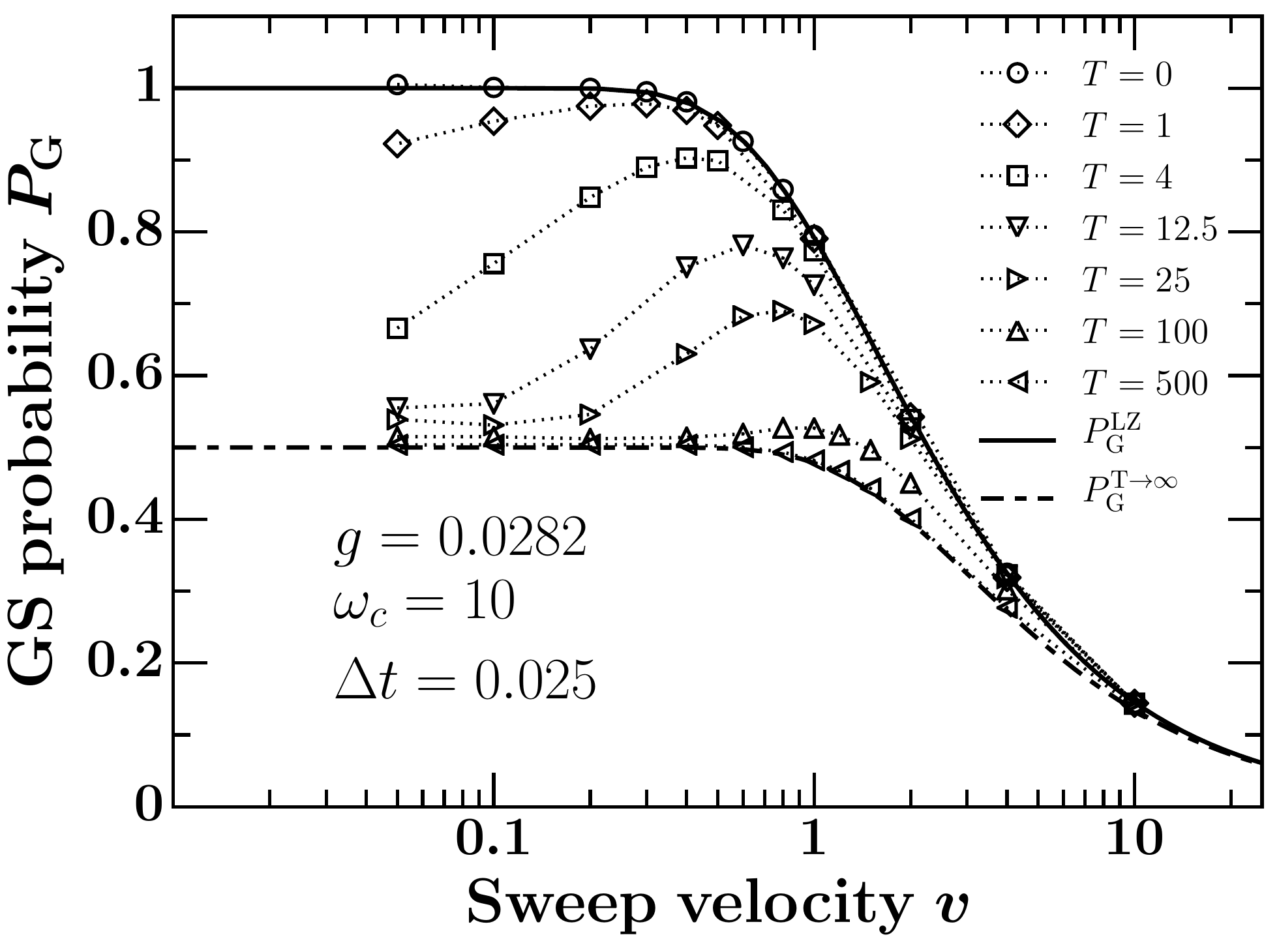}
 \end{center}
 \caption{Grouns state probability as a function
 of the sweep velocity $v$ in the Landau-Zener
 model coupled to a bath. 
 We set the initial time and final time as
 $t_{\rm in} = -40 v$ and $t_{\rm fin} = +40 v$
 and the initial state at the ground state at $t_{\rm in}$.
 Other parameters used in computation are shown in the panel.
 The cutoff in the range of the interaction along the
 time direction and the maximum number of states kept in the
 MPS representation are fixed at $l_c = 40$ and $D_t = 12$.
 $P_{\rm G}^{\rm LZ}$
 and $P_{\rm G}^{T\to\infty}$ denote the ground state
 probability of the isolated Landau-Zener model and
 that of the Landau-Zener model coupled to the
 bath with infinite temperature.
 The results for finite temperatures are obtained
 by the QUAPI-MPS method.
 With increasing the tempereture of the bath,
 the ground state probability decreases from $P_{\rm G}^{\rm LZ}$
 down to $P_{\rm G}^{T\to\infty}$.
 The results for finite temperatures are quantitatively
 consistent with those in Ref. \citen{bib:NalbachPRL2009}.
 }
 \label{fig:LZ}
\end{figure}

In the presence of the bath, however, the analytic solution
is not known in general, except for the zero temperature and
the high temperature limit. In the high temperature limit, 
$P_{\rm G}$ is modified as
$P_{\rm G}^{T\to\infty} = \frac{1}{2}(1 - \exp(-\pi/v))$
\cite{bib:KayanumaJPSJ1984}.
Figure \ref{fig:LZ} shows numerical results using
our QUAPI-MPS method. For performing the simulation,
we fixed parameters as $g = 0.0282$, $\omega_c = 10$,
and $\mathit{\Delta}t = 0.1$. The initial time and final
time are set as $t_{\rm in} = -40v$ and $t_{\rm fin} = +40v$,
respectively. The initial state is set at the ground state
of $H_{\rm LZ}(t_{\rm in})$.
The cutoff in the range of interaction along the time direction
and the maximum number of states kept in the MPS representation are fixed at
$l_c = 40$ and $D_t = 12$, respectively.
The ground state probability decreases with increasing the
temperature from $P_{\rm G}^{\rm LZ}$ down to $P_{\rm G}^{T\to\infty}$.
A similar calculation has been done by Nalbach and
Thorwart\cite{bib:NalbachPRL2009}, who implemented QUAPI without MPS. 
Our results are quantitatively consistent with their results.

\subsection{Test on eight spins}
\begin{figure}[t]
\includegraphics[width=8cm, bb = 22 9 775 578]{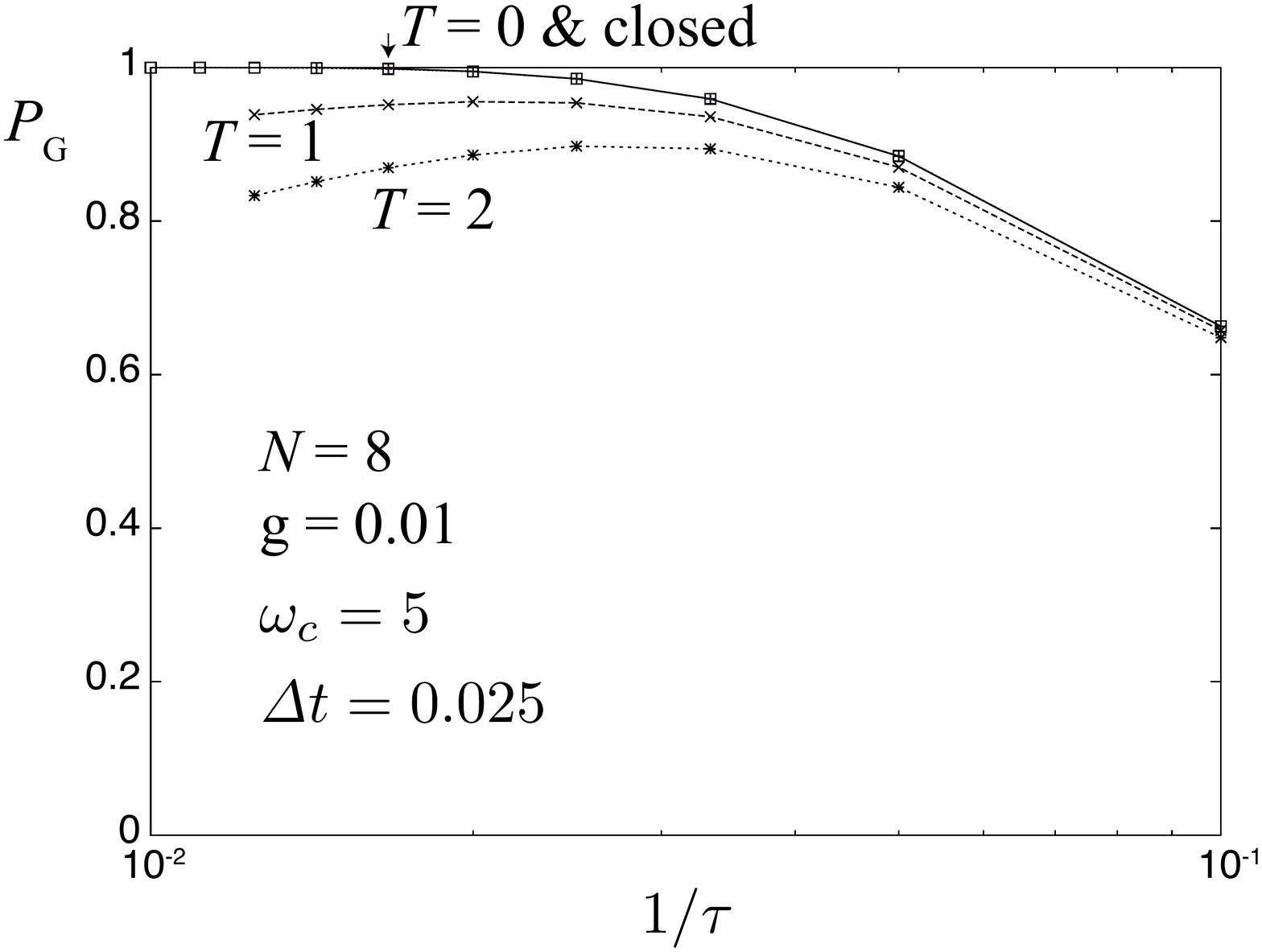}
 \caption{Ground-state probability at the final time $t = \tau$.
 Parameters are fixed as $N = 8$, $g = 0.01$, $\omega_c = 5$, 
 $\mathit{\Delta}t = 0.025$, and $l_c = 100$.
 The numbers of states in the MPS representation are chosen
 as $D_t = 20$ and $D_s \leq 80$.
 Results for the closed system were obtained by solving
 the time-dependent Bogoliubov-de Gennes equation for
 the equivalent free fermion model.
 The result at $T = 0$ is identical to that for the closed
 system (i.e., $g = 0$). With increasing the temperature, the ground-state
 probability decreases. When $T > 0$, the
 ground-state probability increases with increasing $\tau$ for small $\tau$,
 and decreases with $\tau$ for large $\tau$.
 } 
\label{fig:N8}
\end{figure}

We next consider quantum annealing of the pure Ising chain
with eight spins. The Hamiltonian is given by Eq.~(\ref{eq:H_PureIsing}).
Figure \ref{fig:N8} shows the ground-state probability
$P_{\rm G}$
that the ground states are found in the final state at $t = \tau$.
We fixed $\omega_c = 5$, $\mathit{\Delta}t = 0.025$, 
$g = 0.01$, and $l_c = 100$. The numbers of states kept fot
the MPS representation are set as $D_t = 20$ and $D_s = 80$.
When $T = 0$, $P_{\rm G}$
is almost identical with the values for the closed system ($g = 0$).
When $T > 0$, $P_{\rm G}$ is lower than $T = 0$, and increases
with increasing $\tau$ for small $\tau$ but turns into a decrease
for large $\tau$.
This behavior of $P_{\rm G}$ with respect to $\tau$ for finite
temperatures is universal as shown above for larger systems.
This is understood as follows. When $\tau$ is small, 
quantum annealing ends before the system is influenced by the bath.
However, when $\tau$ is large, the influence by the
bath is strong so that $P_{\rm G}$ is lowered.

\begin{figure}[t]
\includegraphics[width=8cm, bb = 0 0 842 595]{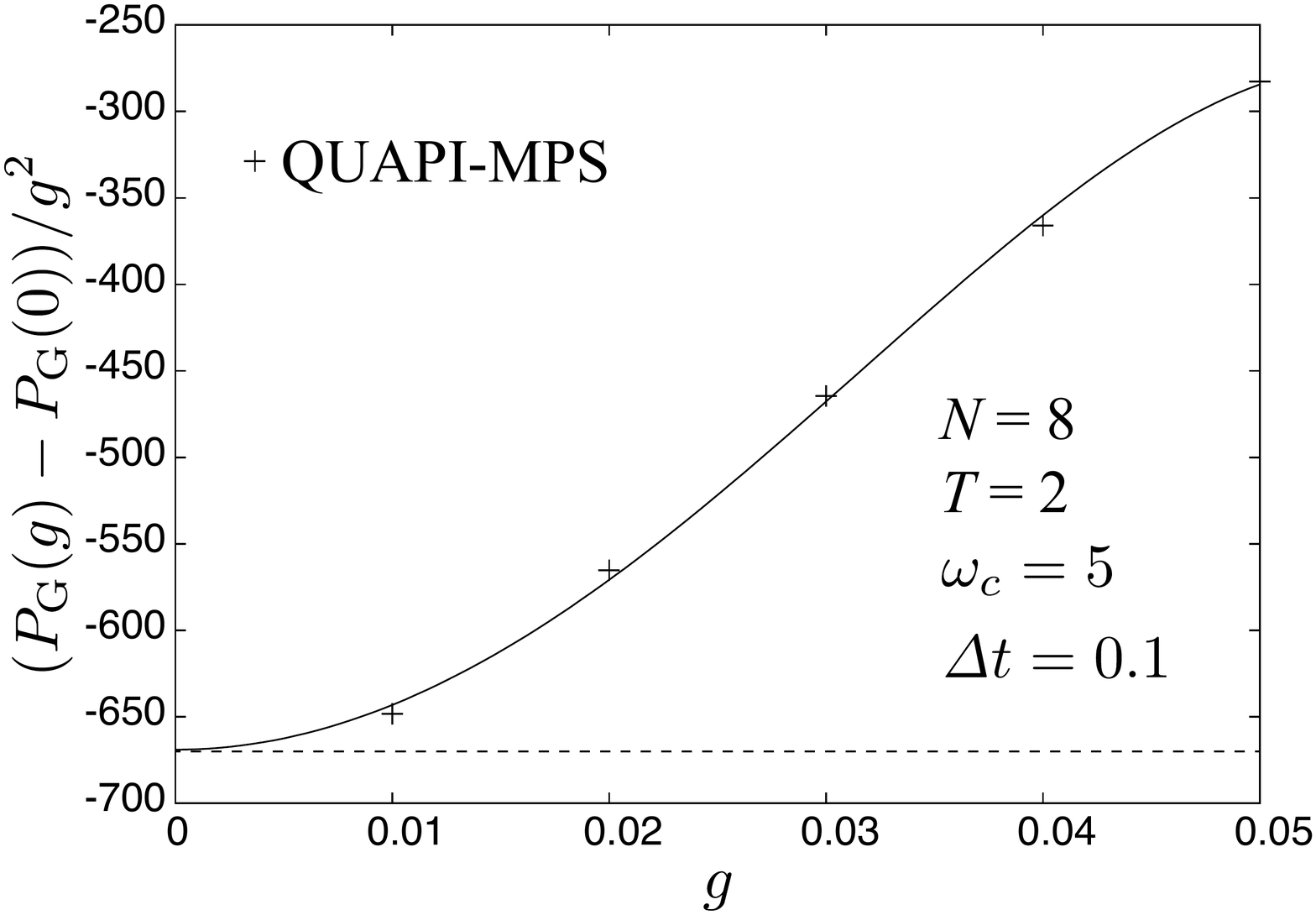}
 \caption{Deviation of the ground-state probability from the value for
 the closed system ($g = 0$) as a function of $g$. Symbols are obtained
 by QUAPI-MPS method. The solid curve represents a fit of QUAPI-MPS data
 by $a + bg^2 + cg^4$. The intercept is estimated as $a = 699\pm 7$.
 The dassed line represents $P_2 = 670$. The extrapolation of QUAPI-MPS
 data to $g = 0$ provides an excellent agreement with the perturbative
 calculation. Both methods share parameters $N = 8$, $T = 2$,
 $\omega_c = 5$, and $\tau = 30$.
 We fixed $\mathit{\Delta}t = 0.1$, $l_c = 100$, $D_t = 12$
 and $D_s = 64$ for QUAPI-MPS.
 }
 \label{fig:N8_P2}
\end{figure}

As mentioned in Sec.~1,
the exact computation of the time-dependent reduced density
matrix is too difficult to achieve even in $N=8$ spin systems.
Hence it is difficult to compare results by our method to exact
ones.
Here we focus on the perturbative expansion with respect to
the system-bath coupling $g^2$. 
As shown in Appendix \ref{sec:App:pert},
the ground-state probability
at final time $t = \tau$ is written up to the order of $g^2$ as
\begin{equation}
 P_{\rm G}(g) \approx P_{\rm G}(0) + g^2 P_2 ,
\end{equation}
\begin{eqnarray}
 &&\hspace{-2em}P_2 := \sum_{\Psi_{\rm G}}
  \Biggl[
  \int_{0}^{\tau}\int_{0}^{\tau}dt_2dt_1K(t_2-t_1) \nonumber\\
 &&\times
  \sum_{j=1}^N\langle\Psi_{\rm G}|\mathcal{U}_{\rm S}(\tau)
  \sigma_j^{z{\rm I}}(t_1)|\Psi_{\rm in}\rangle\langle\Psi_{\rm in}|
  \sigma_j^{z{\rm I}}(t_2)\mathcal{U}_{\rm S}^{\dagger}(\tau)
  |\Psi_{\rm G}\rangle \nonumber\\
 &&- \int_0^{\tau}dt_2\int_0^{t_2}dt_1
  \Bigl(
   K(t_2-t_1) \nonumber\\
 &&\times\sum_{j=1}^N\langle\Psi_{\rm G}|\mathcal{U}_{\rm S}(\tau)
   \sigma_j^{z{\rm I}}(t_2)\sigma_j^{z{\rm I}}(t_1)
   |\Psi_{\rm in}\rangle\langle\Psi_{\rm in}|
   \mathcal{U}_{\rm S}^{\dagger}|\Psi_{\rm G}\rangle
   + \mbox{c.c.}
  \Bigr)
  \Biggr], \nonumber\\
 &&
\end{eqnarray}
where $\sigma_j^{z{\rm I}}(t) :=
\mathcal{U}_{\rm S}^{\dagger}(t)\sigma_j^z\mathcal{U}_{\rm S}(t)$
is the interaction picture of $\sigma_j^z$,
and the summation with respect to $\Psi_{\rm G}$ implies the
summation over the two fully polarized ground states of
$H_{\rm S}(\tau)$. When $N = 8$, the quantity $P_2$ can be accurately computed
by solving the Schr\"odinger equation directly. For instance,
we estimated $P_2 = -670.30\pm 0.02$ for $T = 2$ and $\omega_c = 5$,
by discretizing the integrals over
$t_1$ and $t_2$ and extrapolating the data to the continuous limit.
Figure \ref{fig:N8_P2} shows the comparison between QUAPI-MPS
and the perturbative expansion.
For QUAPI-MPS, we chose $\mathit{\Delta}t = 0.1$, $l_c = 100$,
$D_t = 12$ and $D_s = 64$.
One can see that the extrapolated value of
$(P_{\rm G}(g) - P_{\rm G}(0))/g^2$ by QUAPI-MPS to $g = 0$
agrees excellently with $P_2$ obtained by the perturbative calculation.

\section{Concluding remarks}\label{sec:Conclusion}

We showed numerical results on quantum annealing of
pure and random Ising chains coupled to bosonic baths.
When the system-bath coupling is weak ($g = 0.01$),
the baths with zero temperature hardly influences
quantum annealing. However, even if the temperature of the
bath is zero, the bath hinders quantum annealing 
with strengthening the system-bath coupling at least in the
pure system.
This result is nontrivial bacause the dissipation due to
a sufficiently low temperature may suppress the nonadiabatic
excitation during the time evolution. Our result implies that
this does not happen.
Although the kink density is larger than the closed situation,
it decreases monotonically with increasing the annealing time $\tau$
even though the system-bath coupling is not weak whenever the
temperature is zero. This does not mean, however, that
one can get the solution with probability one by infinitely slow quantum
annealing. This is because, when the system-bath coupling is not weak,
the system does not necessarily equilibriate at its ground state
even if the bath is at the ground state initially.
Such a picture is different when the temperature is finite.
When $T > 0$, the kink density first decays and then turns to
an increase with increasing the annealing time $\tau$.
This is a universal feature for pure and random systems.

In the present paper, we gave an explanation of our QUAPI-MPS method
for the numerical computation of the reduced density matrix.
Advantages of the present method are as follows.
(i) It does not rely on the Born-Markov approximation. Therefore
one can apply it to a strong system-bath coupling.
(ii) The approximations used in the method, the Trotter decomposition,
the limited matrix dimension, and the cutoff in the interaction
range along the time direction, is controllable in the sense that
one can improve these approximations by changing parameters.
(iii) The accessible system size is up to $N\sim 10^2$. The complexity
of the present method scales as a polynomial in $N$ and $M$, where $M$
 stands for the Trotter number.
 (iv) One can engage the infinite size scheme in present method. Then one
 can apply the QUAPI-MPS method to the infinite size, if the system is homogeneous.
 (v) The extention of the present method to other type of spin-spin
 interactions as well as spin-boson interactions is possible. 
 The disadvantage of the present method, on the other hand, is that
 it should not work if a kind of the entanglement entropy
 computed by the (square of) singular values obeys so-called the volume
 law.
 We are not sure so far when this happens, though we have not
 encountered this in studying quantum annealing
 of Ising chains.
 It is an important open issue
 to determine the limitation of the present method. 
 
The study of the time evolution of an open quantum many-body system is
important with no doubt in the development of near-term quantum
computers. In order to confirm the correctness of the operation, the
comparison of experimental results with numerical calculation is
necessary. Moreover, numerical study may provide suggestions on
designing a novel quantum computation, such as a
dissipation assisted quantum computation \cite{bib:SmelyanskiyPRL2017}.
The present method will be
useful for many such studies.  
Apart from the issue of quantum computer, a lot of interesting problems
remain to be solved regarding the time evolution of an open quantum
many-body system. In the present paper, we have not discussed scaling
properties of the kink density, that are associated with the
Kibble-Zurek mechanism. The modification of the
Kibble-Zurek scaling in an open system is an urgent issue. Another
problem is the relaxation or thermalization of a quantum many-body
system coupled to an environment.
It is interesting to ask what the steady state
or the equilibrium state of an open quantum system is in the presence of
driving, such as a quantum quench. We believe that our method should
mark the beginning of a new era in the study of open quantum many-body
systems.

\begin{acknowledgment}
\acknowledgment
 One of the authors (S.S.) acknowledges fruitful discussions with
 L. Arceci, S. Barbarino, and G. E. Santoro. The present work was
 partly supported by JSPS KAKENHI, Japan through Grant No. 26400402.
\end{acknowledgment}

\appendix
\section{Path integral}\label{sec:App:PI}

In this appendix, we derive the QUAPI formula, Eq.~(\ref{eq:rho_S_QUAPI}),
for the reduced density matrix.

The reduced density operator $\rho_{\rm S}(t)$ is written using
the time-evolution operators as
\begin{equation}
 \rho_{\rm S}(t) = \mathcal{U}_{\rm S}(t)
  {\rm Tr}_{\rm B}
  \left(
   \mathcal{U}_{\rm int}(t)\rho_{\rm in}\mathcal{U}_{\rm
   int}^{\dagger}(t)\right)
  \mathcal{U}_{\rm S}^{\dagger}(t) .
\end{equation}
Applying the Trotter decomposition,
$\mathcal{U}_{\rm int}(t = M\mathit{\Delta}t)$ can be written
up to the order of $(\mathit{\Delta}t)^2$ as
\begin{eqnarray}
 &&\hspace{-2em}\mathcal{U}_{\rm int}(M\mathit{\Delta}t)\\
  &&\cong
  e^{-i H_{\rm int}^{\rm I}(M\mathit{\Delta} t)\frac{\mathit{\Delta} t}{2}}
  e^{-i H_{\rm int}^{\rm I}((M-1)\mathit{\Delta} t)\mathit{\Delta} t}
  \cdots
  e^{-i H_{\rm int}^{\rm I}(\mathit{\Delta} t)\mathit{\Delta} t}
  e^{-i H_{\rm int}^{\rm I}(0)\frac{\mathit{\Delta} t}{2}} ,
  \nonumber
\end{eqnarray}
and each exponential operator is written as
\begin{eqnarray}
 &&\hspace{-2em}e^{-iH_{\rm int}^{\rm I}(l\mathit{\Delta} t)\mathit{\Delta}t}\nonumber\\
 &&\hspace{-2em}=
  \mathcal{U}_{\rm S}^{\dagger}(l\mathit{\Delta} t)\left[\prod_{j,a}
   e^{-i\mathit{\Delta} t\sigma_j^z\lambda_{ja}b_{ja}^{\dagger}e^{i\omega_a
   l\mathit{\Delta} t}}
   e^{-i\mathit{\Delta} t\sigma_j^z\lambda_{ja}b_{ja}e^{-i\omega_a l\mathit{\Delta} t}}
   e^{-\frac{1}{2}\mathit{\Delta} t^2\lambda_{ja}^2}
       \right] \nonumber\\
 &&\times \mathcal{U}_{\rm S}(l\mathit{\Delta} t) .
\label{eq:App:exponentialOp}
\end{eqnarray}
See Eqs.~(\ref{eq:Trotter}) and (\ref{eq:exponentialOp}).

Now we introduce the coherent state $|z\}$
of the boson operator $b_{ja}$ in such a way that
\begin{equation}
 b_{ja}|z\} = z_{ja}|z\} ~~~ \forall j,a ,
\end{equation}
\begin{equation}
 \{ z|z'\} = \prod_{j,a}e^{z_{ja}^{\ast}z_{ja}'} .
\end{equation}
Using the notation for the basis
$|\boldsymbol{\sigma}\rangle$
of the spin state such that
\begin{equation}
 \sigma_j^z|\boldsymbol{\sigma}\rangle
 = \sigma_j|\boldsymbol{\sigma}\rangle ,
\end{equation}
the completeness relation is given by
\begin{equation}
 1 = \sum_{\boldsymbol{\sigma}}
  |\boldsymbol{\sigma}\rangle\langle\boldsymbol{\sigma}|\otimes
  \int\prod_{j,a}\mathcal{D}z_{ja} e^{-z_{ja}^{\ast}z_{ja}}|z\}\{z| ,
\end{equation}
where the summation is taken over $\sigma_1,\cdots,\sigma_N$
and $\mathcal{D}z_{ja}:=(2\pi)^{-1}dz_{ja}dz_{ja}^{\ast}$.
Inserting the completeness relation between the right square
bracket and $\mathcal{U}_{\rm S}(l\mathit{\Delta}t)$
in Eq.~(\ref{eq:App:exponentialOp}),
one obtains the path-integral representation of
$\rho_{\rm S}(M\mathit{\Delta}t)$ as follows:
\begin{eqnarray}
 &&\hspace{-2em}\langle\boldsymbol{\sigma}_{M}|\rho_{\rm
  S}(M\mathit{\Delta}t) |\boldsymbol{\tau}_{M}\rangle
  \cong
  \sum_{\boldsymbol{\sigma}_0}\sum_{\boldsymbol{\sigma}_1}
  \cdots\sum_{\boldsymbol{\sigma}_{M-1}}
  \sum_{\boldsymbol{\tau}_0}\sum_{\boldsymbol{\tau}_1}
  \cdots\sum_{\boldsymbol{\tau}_{M-1}}\nonumber\\
 &&\times \langle\boldsymbol{\sigma}_M|
  \mathcal{U}_{\rm S}(M\mathit{\Delta}t)
  \mathcal{U}_{\rm S}^{\dagger}((M-t)\mathit{\Delta}t)
  |\boldsymbol{\sigma}_{M-1}\rangle\times\cdots \nonumber\\
 &&  \times\langle\boldsymbol{\sigma}_1|
  \mathcal{U}_{\rm S}(\mathit{\Delta}t)\mathcal{U}_{\rm S}^{\dagger}(0)
  |\boldsymbol{\sigma}_0\rangle
  \langle\boldsymbol{\sigma}_0|\Psi_{\rm in}\rangle \nonumber\\
 &&\times\langle\Psi_{\rm in}|\boldsymbol{\tau}_0\rangle
  \langle\boldsymbol{\tau}_0|
  \mathcal{U}_{\rm S}(0)\mathcal{U}_{\rm S}^{\dagger}(\mathit{\Delta}t)
  |\boldsymbol{\tau}_1\rangle\times\cdots \nonumber\\
 &&\times  \langle\boldsymbol{\tau}_{M-1}|
  \mathcal{U}_{\rm S}((M-1)\mathit{\Delta}t)
  \mathcal{U}_{\rm S}^{\dagger}(M\mathit{\Delta}t)
  |\boldsymbol{\tau}_M\rangle \nonumber\\
 &&\times\prod_{j,a}
  \int\mathcal{D}z_{ja}\prod_{l=0}^M\mathcal{D}z_{ja}^{(l)}
  \mathcal{D}w_{ja}^{(l)}\nonumber\\
 &&\times \exp\Biggl[
  - \left((M-1)+\frac{1}{2}\right)\mathit{\Delta}t^2\lambda_{ja}^2
  - \boldsymbol{z}^{\dagger}A\boldsymbol{z}
  - \boldsymbol{z}^{\dagger}\boldsymbol{x}_{\sigma}
  - \boldsymbol{y}_{\sigma}^{\dagger}\boldsymbol{z}
  \nonumber\\
 && - \boldsymbol{w}^{\dagger}A\boldsymbol{w}
  - \boldsymbol{w}^{\dagger}\boldsymbol{y}_{\tau}
  - \boldsymbol{x}'^{\dagger}\boldsymbol{w}
  \nonumber\\
 && - z_{ja}^{\ast}z_{ja} - z_{ja}^{(0)\ast}z_{ja}^{(0)}
  - w_{ja}^{(0)\ast}w_{ja}^{(0)}
  + z_{ja}^{(0)\ast}w_{ja}^{(0)}e^{-\beta\omega_{ja}}
  \nonumber\\
 && -
  i\frac{\mathit{\Delta}t}{2}\lambda_{ja}e^{i\omega_a M\mathit{\Delta}t}
  \sigma_{jM}z_{ja}^{\ast}
  - i\frac{\mathit{\Delta}t}{2}\lambda_{ja}\sigma_{j0}z_{ja}^{(0)}
  \nonumber\\
 && +
  i\frac{\mathit{\Delta}t}{2}\lambda_{ja}e^{-i\omega_a
  M\mathit{\Delta}t}
  \tau_{jM}z_{ja}
  + i\frac{\mathit{\Delta}t}{2}\lambda_{ja}\tau_{j0}w_{ja}^{(0)}
\Biggr] ,
\end{eqnarray}
where we defined
\begin{equation}
 \boldsymbol{z} =
  \left(
   \begin{array}{c}
    z_{ja}^{(M)}\\
    \vdots\\
    z_{ja}^{(1)}
   \end{array}
	      \right) ,~~~
 \boldsymbol{w} =
  \left(
   \begin{array}{c}
    w_{ja}^{(M)}\\
    \vdots\\
    w_{ja}^{(1)}
   \end{array}
	      \right) ,~~~
\end{equation}
\begin{equation}
  A =
  \left(
   \begin{array}{ccccc}
    1 & -1 & & & \\
    & 1 & -1 & & \\
    & & 1 & & \\
    & & & \ddots & \\
    & & & & -1 \\
    & & & & 1
   \end{array}
  \right) ,
\end{equation}
\begin{equation}
 \boldsymbol{x}_{\sigma}
  = \left(
     \begin{array}{c}
      i\mathit{\Delta}t\lambda_{ja}e^{i\omega_{ja}(M-1)\mathit{\Delta}t}\sigma_{jM-1}\\
      \vdots \\
      i\mathit{\Delta}t\lambda_{ja}e^{i\omega_{ja}\mathit{\Delta}t}\sigma_{j1}\\
      i\frac{\mathit{\Delta}t}{2}\lambda_{ja}\sigma_{j0} - z_{ja}^{(0)}
     \end{array}
    \right) ,
\end{equation}
\begin{equation}
  \boldsymbol{y}_{\sigma}
  = \left(
     \begin{array}{c}
      -i\frac{\mathit{\Delta}t}{2}\lambda_{ja}e^{i\omega_{ja}M\mathit{\Delta}t}\sigma_{jM}
       - z_{ja}\\
      -i\mathit{\Delta}t\lambda_{ja}e^{i\omega_{ja}(M-1)\mathit{\Delta}t}\sigma_{jM-1}\\
      \vdots\\
      -i\mathit{\Delta}e^{i\omega_{ja}\mathit{\Delta}t}\sigma_{j1}
     \end{array}
    \right) ,
\end{equation}
and
\begin{equation}
 \boldsymbol{x}'_{\tau}
  = \left(
     \begin{array}{c}
      i\mathit{\Delta}t\lambda_{ja}e^{i\omega_{ja}(M-1)\mathit{\Delta}t}\tau_{jM-1}\\
      \vdots \\
      i\mathit{\Delta}t\lambda_{ja}e^{i\omega_{ja}\mathit{\Delta}t}\tau_{j1}\\
      i\frac{\mathit{\Delta}t}{2}\lambda_{ja}\tau_{j0} - w_{ja}^{(0)}
     \end{array}
    \right) .
\end{equation}
Since the integrals over $z$'s and $w$'s are Gaussian, they are
performed analytically. The result is given by
\begin{eqnarray}
 &&\hspace{-2em}\langle\boldsymbol{\sigma}_{M}|\rho_{\rm S}(M\mathit{\Delta}t)
  |\boldsymbol{\tau}_M\rangle
  \cong
  \sum_{\boldsymbol{\sigma}_0}\sum_{\boldsymbol{\sigma}_1}
  \cdots\sum_{\boldsymbol{\sigma}_{M-1}}
  \sum_{\boldsymbol{\tau}_0}\sum_{\boldsymbol{\tau}_1}
  \cdots\sum_{\boldsymbol{\tau}_{M-1}}\nonumber\\
 &&\times \langle\boldsymbol{\sigma}_M|
  \mathcal{U}_{\rm S}(M\mathit{\Delta}t)
  \mathcal{U}_{\rm S}^{\dagger}((M-t)\mathit{\Delta}t)
  |\boldsymbol{\sigma}_{M-1}\rangle\times\cdots \nonumber\\
 &&  \times\langle\boldsymbol{\sigma}_1|
  \mathcal{U}_{\rm S}(\mathit{\Delta}t)\mathcal{U}_{\rm S}^{\dagger}(0)
  |\boldsymbol{\sigma}_0\rangle
  \langle\boldsymbol{\sigma}_0|\Psi_{\rm in}\rangle \nonumber\\
 &&\times\langle\Psi_{\rm in}|\boldsymbol{\tau}_0\rangle
  \langle\boldsymbol{\tau}_0|
  \mathcal{U}_{\rm S}(0)\mathcal{U}_{\rm S}^{\dagger}(\mathit{\Delta}t)
  |\boldsymbol{\tau}_1\rangle\times\cdots \nonumber\\
 &&\times  \langle\boldsymbol{\tau}_{M-1}|
  \mathcal{U}_{\rm S}((M-1)\mathit{\Delta}t)
  \mathcal{U}_{\rm S}^{\dagger}(M\mathit{\Delta}t)
  |\boldsymbol{\tau}_M\rangle \nonumber\\
 &&\prod_{j=1}^N
  \exp\Biggl[
  - \left(M-1+ \frac{1}{2}\right)\mathit{\Delta}^2 L
  + \mathit{\Delta}t^2L\sum_{l=0}^M\sigma_{j,l}\tau_{j,l}
  \nonumber\\
 && - \mathit{\Delta}t^2\sum_{M\geq l > m\geq 0}
  K((l-m)\mathit{\Delta}t)\sigma_{j,l}\sigma_{j,m}
  \nonumber\\
 && - \mathit{\Delta}t^2\sum_{M\geq l > m\geq 0}
  K^{\ast}((l-m)\mathit{\Delta}t)\tau_{j,l}\tau_{j,m}
  \nonumber\\
 && + \mathit{\Delta}t^2\sum_{M\geq l > m\geq 0}
  K^{\ast}((l-m)\mathit{\Delta}t)\sigma_{j,l}\tau_{j,m}
  \nonumber\\
 && + \mathit{\Delta}t^2\sum_{M\geq l > m\geq 0}
  K((l-m)\mathit{\Delta}t)\tau_{j,l}\sigma_{j,m}
  \Biggr] ,
  \label{eq:App:rho_S_1}  
\end{eqnarray}
where $L$ and $K(t)$ are defined by Eqs.~(\ref{eq:L}) and
(\ref{eq:Kernel}), respectively.
We remark that $\sigma_{j,0}$, $\sigma_{j,M}$, $\tau_{j,0}$, and
$\tau_{j,M}$ with $j = 1,\cdots, N$ must be multiplied by
the factor $\frac{1}{2}$ in the above equation.

We move on to the spin degree of freedom. One can notice that
the product
$\mathcal{U}_{\rm S}(l\mathit{\Delta}t)
\mathcal{U}_{\rm S}^{\dagger}((l-1)\mathit{\Delta}t)$
is the time evolution operator from $t = (l-1)\mathit{\Delta}t$
to $t = l\mathit{\Delta}t$. Applying the symmmetric
decomposition for the exponential operator,
one obtains
\begin{equation}
 \mathcal{U}_{\rm S}(l\mathit{\Delta}t)
  \mathcal{U}_{\rm S}^{\dagger}((l-1)\mathit{\Delta}t)
  =
  e^{-iH_{\rm S}(l\mathit{\Delta}t)\frac{\mathit{\Delta}t}{2}}
  e^{-iH_{\rm S}((l-1)\mathit{\Delta}t)\frac{\mathit{\Delta}t}{2}}
  + \mathcal{O}(\mathit{\Delta}t^3) .
\label{eq:App:U1}
\end{equation}
Using the notation $H_l^z = - \sum_{j=1}^{N-1}J_j(l\mathit{\Delta}t)\sigma_j^z\sigma_{j+1}^z$
and $H_l^x = - \sum_{j=1}^Nh(l\mathit{\Delta}t)\sigma_j^x$,
an exponential operator can be further decomposed as
\begin{equation}
 e^{-H_{\rm S}(l\mathit{\Delta}t)\frac{\mathit{\Delta}t}{2}}
  =
  e^{-iH_l^z\frac{\mathit{\Delta}t}{4}}e^{-iH_l^x\frac{\mathit{\Delta}t}{2}}
  e^{-iH_l^z\frac{\mathit{\Delta}t}{2}} +
  \mathcal{O}(\mathit{\Delta}t^3) .
\end{equation}
Therefore Eq.~(\ref{eq:App:U1}) can be arranged into
\begin{eqnarray}
 &&\hspace{-2em} \mathcal{U}_{\rm S}(l\mathit{\Delta}t)
  \mathcal{U}_{\rm S}^{\dagger}((l-1)\mathit{\Delta}t)
  \nonumber\\
 &&\cong
  e^{-\frac{i}{4}H_l^z\mathit{\Delta}t}
  e^{-\frac{i}{2}H_l^x\mathit{\Delta}t}
  e^{-\frac{i}{4}H_l^z\mathit{\Delta}t}
  e^{-\frac{i}{4}H_{l-1}^z\mathit{\Delta}t}
  e^{-\frac{i}{2}H_{l-1}^x\mathit{\Delta}t}
  e^{-\frac{i}{4}H_{l-1}^z\mathit{\Delta}t} \nonumber\\
 &&\cong
  e^{-\frac{i}{8}(3H_l^z+ H_{l-1}^z)\mathit{\Delta}t}
  e^{-\frac{i}{2}(H_l^x + H_{l-1}^x)\mathit{\Delta}t}
  e^{-\frac{i}{8}(H_l^z+ 3H_{l-1}^z)\mathit{\Delta}t}
\label{eq:App:U2}
\end{eqnarray}
up to the order $\mathit{\Delta}t^2$, where we note 
\begin{eqnarray}
 &&\hspace{-2em}e^{-\frac{i}{2}H_l^x\mathit{\Delta}t}
  e^{-\frac{i}{4}(H_l^z + H_{l-1}^z)\mathit{\Delta}t}
  e^{-\frac{i}{2}H_{l-1}^x\mathit{\Delta}t}
  \\
 && = e^{-\frac{i}{8}(H_l^z + H_{l-1}^z)\mathit{\Delta}t}
  e^{-\frac{i}{2}(H_l^x + H_{l-1}^x)\mathit{\Delta}t}
  e^{-\frac{i}{8}(H_l^z + H_{l-1}^z)\mathit{\Delta}t}
  + \mathcal{O}(\mathit{\Delta}t^3) .
  \nonumber
\end{eqnarray}
The matrix element of Eq.~(\ref{eq:App:U2}) is written as
\begin{eqnarray}
 &&\hspace{-2em}\langle\boldsymbol{\sigma}_l|
  \mathcal{U}_{\rm S}(l\mathit{\Delta}t)
  \mathcal{U}_{\rm S}((l-1)\mathit{\Delta}t)
  |\boldsymbol{\sigma}_{l-1}\rangle
  \nonumber\\
 &&\cong
  e^{-\frac{i}{8}(3H_l^z(\boldsymbol{\sigma}_l)
  + H_{l-1}^z(\boldsymbol{\sigma}_l))\mathit{\Delta}t}
  \langle\boldsymbol{\sigma}_l|
  e^{-\frac{i}{2}(H_l^x + H_{l-1}^x)\mathit{\Delta}t}
  |\boldsymbol{\sigma}_{l-1}\rangle \nonumber\\
 &&\hspace{1em}\times
  e^{-\frac{i}{8}(H_l^z(\boldsymbol{\sigma}_{l-1})
  + 3 H_{l-1}^z(\boldsymbol{\sigma}_{l-1}))}
  \nonumber\\
 &&= \prod_{j=1}^N
  \left(\frac{i}{2}\sin 2
   \frac{h(l\mathit{\Delta}t) +
   h((l-1)\mathit{\Delta}t)}{2}\mathit{\Delta}t
				  \right)^{\frac{1}{2}}\nonumber\\
 &&\hspace{1em}\times
  \exp\Biggl[
  - \frac{i}{8}\left(3 H_l^z(\boldsymbol{\sigma}_l) +
	       H_{l-1}^z(\boldsymbol{\sigma}_l)\mathit{\Delta}t\right)
  \nonumber\\
 &&\hspace{1em} - \frac{i}{8}\left(H_l^z(\boldsymbol{\sigma}_{l-1}) +
	       H_{l-1}^z(\boldsymbol{\sigma}_{l-1})\mathit{\Delta}t\right)
 + \sum_{j=1}^{N}\gamma_l\sigma_{jl}\sigma_{jl-1}
 \Biggr] ,\nonumber\\
\label{eq:App:U3}
\end{eqnarray}
where
$H_l^z(\boldsymbol{\sigma}_m) = - \sum_{j=1}^{N-1}
J_{j}(l\mathit{\Delta}t)\sigma_{j,m}\sigma_{j+1,m}$
and $\gamma_l$ is defined by Eq.~(\ref{eq:gamma_l}).
Substituting Eq.~(\ref{eq:App:U3}) for the matrix
elements in Eq.~(\ref{eq:App:rho_S_1}), one obtains
Eq.~(\ref{eq:rho_S_QUAPI}).

\section{Perturbation expansion of the reduced density matrix}\label{sec:App:pert}

In this appendix, we describe the perturbation expantion of the
reduced density matrix up to the second order with respect to
the system-bath coupling.

As mentioned in Sec.~\ref{sec:Model}, we decompose the
time evolution operator $\mathcal{U}(t)$ as
$\mathcal{U}(t) = \mathcal{U}_{\rm S}(t)\mathcal{U}_{\rm
B}(t)\mathcal{U}_{\rm int}(t)$, where $\mathcal{U}_{\rm S}(t)$
and $\mathcal{U}_{\rm B}(t)$ are the time-evolution operatores
for the isolated system and the bath, respectively.
Then $\mathcal{U}_{\rm int}(t)$ obeys an equation
\begin{equation}
 i\frac{d}{dt}\mathcal{U}_{\rm int}(t) = H^{\rm I}_{\rm int}(t)
  \mathcal{U}_{\rm int}(t) ,
\end{equation}
where
$H^{\rm I}_{\rm int}(t) =
\mathcal{U}_{\rm B}^{\dagger}(t)\mathcal{U}_{\rm
S}^{\dagger}(t)H_{\rm int}\mathcal{U}_{\rm S}(t)\mathcal{U}_{\rm B}(t)$
is the interaction picture of $H_{\rm int}$.
We hereafter assume $\mathcal{U}_{\rm S}(0)=\mathcal{U}_{\rm
B}(0)=\mathcal{U}_{\rm int}(0)=1$.
Now we consider the perturbation expansion of $\mathcal{U}_{\rm int}$
up to the second order in $H_{\rm int}$ as
\begin{equation}
 \mathcal{U}_{\rm int}(t) \approx 1 + \mathcal{U}^{(1)}_{\rm int}(t)
  + \mathcal{U}^{(2)}_{\rm int}(t) ,
\label{eq:App:U_int_expansion}
\end{equation}
\begin{equation}
 \mathcal{U}^{(1)}_{\rm int}(t)
  = -i\int_0^t dt_1 H^{\rm I}_{\rm int}(t_1) ,
\label{eq:App:U_int_expansion_1}
\end{equation}
\begin{equation}
 \mathcal{U}^{(2)}_{\rm int}(t)
  = (-i)^2\int_0^t dt_2\int_0^{t_2} dt_1 H^{\rm I}_{\rm int}(t_2)
  H^{\rm I}_{\rm int}(t_1) .
\label{eq:App:U_int_expansion_2}
\end{equation}

The density matrix is defined by
$\rho(t) = \mathcal{U}(t)\rho_{\rm in}\mathcal{U}^{\dagger}(t)$.
We assume the initial condition: $\rho(0)=\rho_{\rm in}$
and $\rho_{\rm in}=|\Psi_{\rm in}\rangle\langle\Psi_{\rm in}|\otimes
e^{-\beta H_{\rm B}}/Z_{\rm B}$, where
$|\Psi_{\rm in}\rangle$ denotes the initial state of the system,
$\beta$ is
the inverse temperature of the bath, 
$H_{\rm B}$ is the Hamiltonian of the bath,
and
$Z_{\rm B} = {\rm Tr}_{\rm B}e^{-\beta H_{\rm B}}$.
Using
Eqs.~(\ref{eq:App:U_int_expansion})-(\ref{eq:App:U_int_expansion_2}),
the density matrix is expanded into the perturbation series up to the
second order as 
\begin{eqnarray}
 &&\hspace{-2em}\rho(t) \approx
  \mathcal{U}_{\rm S}(t)\mathcal{U}_{\rm B}(t)\rho_{\rm in}
  \mathcal{U}_{\rm B}^{\dagger}(t)\mathcal{U}_{\rm S}^{\dagger}(t)
  \nonumber  \\
 &&+ \mathcal{U}_{\rm S}(t)\mathcal{U}_{\rm B}(t)
  \left(\mathcal{U}^{(1)}_{\rm int}(t)\rho_{\rm in} +
   \rho_{\rm int}\mathcal{U}^{(1)\dagger}_{\rm int}(t)\right)
  \mathcal{U}_{\rm S}^{\dagger}(t)\mathcal{U}_{\rm B}^{\dagger}(t)
  \nonumber\\
 &&+ \mathcal{U}_{\rm S}(t)\mathcal{U}_{\rm B}(t)
  \left(\mathcal{U}^{(1)}_{\rm int}(t)\rho_{\rm
   in}\mathcal{U}^{(1)\dagger}_{\rm int} \right. \nonumber \\
 &&\hspace{5em}+ \left.
   \mathcal{U}^{(2)}(t)\rho_{\rm in} +
   \rho_{\rm int}\mathcal{U}^{(2)\dagger}_{\rm int}(t)\right)
 \mathcal{U}_{\rm S}^{\dagger}(t)\mathcal{U}_{\rm B}^{\dagger}(t)
 .
 \nonumber\\
 &&
\end{eqnarray}
The perturbation expansion of the reduced density matrix
$\rho_{\rm S}(t):={\rm Tr}_{\rm B}\rho(t)$ is obtained by
taking the trace with respect to the degree of freedom of the bath.
Now we assume the Hamiltonians $H_{\rm B}$ and $H_{\rm int}$ in
Eqs.~(\ref{eq:H_B}) and (\ref{eq:Hint}), and the Ohmic spectral density
defined by Eqs.~(\ref{eq:SpectralDensity1}) and
(\ref{eq:SpectralDensity2}).
Noting ${\rm Tr}_{\rm B}(\mathcal{U}^{(1)}_{\rm int}\rho_{\rm in})=
{\rm Tr}_{\rm B}(\rho_{\rm in}\mathcal{U}^{(1)}_{\rm int})=0$ and
the kernel function $K(t)$ defined by Eq.~(\ref{eq:Kernel}),
we obtain
\begin{eqnarray}
 &&\hspace{-4em}\rho_{\rm S}(t) \approx
  \mathcal{U}_{\rm S}(t)|\Psi_{\rm in}\rangle
  \langle\Psi_{\rm in}|\mathcal{U}_{\rm S}^{\dagger}(t) \nonumber\\
 &&\hspace{-1em}+ \int_0^{t} dt_2\int_0^{t} dt_1 K(t_2-t_1) \nonumber\\
 &&\times \sum_{j=1}^N\mathcal{U}_{\rm S}(t)\sigma_j^{z{\rm I}}(t_1)
  |\Psi_{\rm in}\rangle\langle\Psi_{\rm in}|\sigma_j^{z{\rm I}}(t_2)
  \mathcal{U}_{\rm S}^{\dagger}(t) \nonumber\\
 &&\hspace{-1em}- \int_0^{t}dt_2\int_0^{t_2}dt_1
  \Bigl(K(t_2-t_1) \nonumber\\
 &&\times \sum_{j=1}^N
  \mathcal{U}_{\rm S}(t)\sigma_j^{z{\rm I}}(t_2)\sigma_j^{z{\rm I}}(t_1)
  |\Psi_{\rm in}\rangle\langle\Psi_{\rm in}|
  \mathcal{U}_{\rm S}^{\dagger}(t) + \mbox{h.c.}
  \Bigr) , \nonumber\\
 &&
\label{eq:App:rho_S_expansion}
\end{eqnarray}
where $\sigma_j^{z{\rm I}}(t) := \mathcal{U}_{\rm
S}^{\dagger}(t)\sigma_j^z\mathcal{U}_{\rm S}(t)$
denotes the interaction picture of $\sigma_j^z$.
We note that 
the matrix elements $\mathcal{U}_{\rm S}(t)$ can be
numerically
computed for small systems with $N$ up to about $N=10$ by
solving the Schr\"odinger equation. Using them, one can
evaluate the matrix elements of the right-hand side
of Eq.~(\ref{eq:App:rho_S_expansion}) through a discrete
approximation on integrals.

\end{document}